\documentclass[12pt]{article}

\font\grande=cmr9.5 scaled \magstep4
\font\medio=cmr9.5 scaled \magstep2
\outer\def\beginsection#1\par{\medbreak\bigskip
      \message{#1}\leftline{\bf#1}\nobreak\medskip
\vskip-\parskip
      \noindent} 
\usepackage{graphicx} 
\begin{document}
\bibliographystyle {unsrt}

\titlepage

\begin{flushright}
CERN-PH-TH/2009-213
\end{flushright}

\vspace{15mm}
\begin{center}
{\grande Electric-magnetic duality}\\
\vspace{0.5cm}
{\grande and the conditions of inflationary magnetogenesis}\\
\vspace{1.5cm}
 Massimo Giovannini 
 \footnote{Electronic address: massimo.giovannini@cern.ch} \\
\vspace{1cm}
{{\sl Department of Physics, 
Theory Division, CERN, 1211 Geneva 23, Switzerland }}\\
\vspace{0.5cm}
{{\sl INFN, Section of Milan-Bicocca, 20126 Milan, Italy}}
\vspace*{2cm}

\end{center}

\vskip 1cm
\centerline{\medio  Abstract}
The magnetogenesis scenarios triggered by the early variation of the gauge coupling are critically analyzed. 
In the absence of sources, it is shown that the electric and magnetic power spectra 
can be explicitly computed by means of electric-magnetic duality transformations. The  remnants of a pre-inflationary expansion and the reheating process break explicitly electric-magnetic duality by inducing Ohmic currents. The generation of large-scale magnetic field and the physical distinction 
between electric and magnetic observables stems, in this class of models,  from the final value reached by the conductivity of the plasma right after inflation. Specific numerical examples are given. The physical requirements of viable  magnetogenesis scenarios are spelled out. 
\noindent

\vspace{5mm}

\newpage

\renewcommand{\theequation}{1.\arabic{equation}}
\setcounter{equation}{0}
\section{Early variation of gauge couplings}

Large-scale magnetic fields were firstly detected in the optical by means of polarimetric observations \cite{detect}.
The observational evidence supported the presumption that the Milky Way, as well other spiral galaxies, possess a 
magnetic field whose associated pressure turned out to 
be comparable with the cosmic ray pressure \cite{fermi}. While at the beginning the opinions 
seemed to be diverging \cite{alv} the hypothesis of large-scale magnetism got progressively firmer.
Sixty years have passed  and the  steady development of 
radio-astronomical and microwave observations confirm today the initial vision \cite{detect,fermi} which
represented, incidentally, one of the first successful interplays between high-energy physics and astronomy.

Since the late eighties there is mounting evidence that nearly all
 gravitationally bound systems at large scales are endowed with a corresponding  
magnetic field. Even if this general lore is not as firm as galactic magnetism, 
the general  evidence of  magnetic fields associated with Abell clusters is becoming 
progressively more compelling (see \cite{mg1} for a general review). The existence 
of large-scale magnetic fields in super-clusters is, at the moment, not yet 
established because of intrinsic observational drawbacks which might become less severe with 
the advent of a new generation of experiments such as the ambitious Square Kilometer Array (SKA) \cite{SKA}. 
If confirmed, the recent Auger observations \cite{auger} might constrain 
the magnetic fields present in a cocoon of $60$--$70$ Mpc.

One of the intriguing questions present from the very beginning \cite{fermi} concerns 
the origin of large-scale magnetism. This question became more acute since all spirals 
have magnetic fields with comparable strengths and similar correlation scales \cite{RB}. 
There are certainly important observational issues related to the latter problem 
and at least some of them could be addressed in the framework 
of the SKA project (see, for instance, third paper in \cite{RB}).
At the same time only galactic observations will not tell us, with reasonable confidence, if large-scale 
magnetism is a phenomenon which is primordial or recent. In the present context primordial means
that the magnetic field had to be already present before photon decoupling. The true problem 
before us today is, therefore, not the endorsement of a particular model. The 
true issue before us today concerns  the presence rather than the absence of large-scale 
magnetic fields before photon decoupling \cite{mga}. 
Are we able to exclude, to a certain confidence level, that pre-decoupling 
 magnetic fields did distort the observed Cosmic Microwave Background (CMB) anisotropies 
 and polarization?
This question motivated a bottom-up approach to magnetized CMB observables. 
Recent results show that large-scale magnetic fields larger than $3.5$ nG are excluded 
to 95 \% C.L. and for magnetic spectral indices $n_{\mathrm{B}} = 1.6_{-0.1}^{0.8}$ (see
\cite{mga} and references therein). While further studies along this direction might either 
confirm or infirm the obtained results, it is important to stress that, absent specific bounds and observations 
on pre-decoupling magnetism, the primordial nature of large-scale magnetic field will just become 
the arena of theoretical discussions. Having said this, it is interesting to speculate on the origin 
of the pre-decoupling magnetism even if the final referee, in this dispute, will not be theoretical opinions  
but the observational evidence.

There are, indeed, various models for the origin of large-scale magnetism based 
on diverse physical pictures \cite{mg1}. The mechanisms based  on the 
parametric amplification of minute fluctuations of a putative Abelian field 
allow to obtain correlation scales which are naturally large.
In the present paper the attention shall be confined on those magnetogenesis 
scenarios which are based, in a way or in another, on the assumption 
that the hypermagnetic and hyperelectric fields are coupled 
to a spectator field whose evolution is dictated by a quasi-de Sitter 
dynamics \cite{variation1}. Recently various aspects of these models have been investigated \cite{variation2}.
A related class of models stipulates that the the 
gauge coupling takes part in the dynamics of the background and this can arise 
either in the context of conventional inflationary models \cite{ratra} or in the context of 
string cosmological scenarios \cite{pbb}.  Only some of the considerations 
developed here are applicable to both classes of models and probably only 
the results of section \ref{sec2} are general enough to embrace all the different possibilities. 

Consider, for simplicity, a spatially flat Friedmann-Robertson-Walker 
geometry whose line element can be written, in the conformal time parametrization as 
\begin{equation}
ds^2 = G_{\mu\nu} dx^{\mu} dx^{\nu} = a^2(\tau) [d\tau^2 - d\vec{x}^2], \qquad 
G_{\mu\nu} = a^2(\tau) \eta_{\mu\nu}
\label{met}
\end{equation}
where $\tau$ denotes the conformal time coordinate and $\eta_{\mu\nu}$ is the Minkowsky metric. In what 
follows $\nabla_{\alpha}$ will denote the covariant derivative with respect to the metric (\ref{met}). In the absence of any concentration of particles charged under a putative (Abelian) gauge field the action can be written as
\begin{equation}
S_{Y} = - \frac{1}{16\pi} \int d^{4} x \sqrt{- G}\biggl[\lambda(\psi)  Y_{\alpha\beta} \,Y^{\alpha\beta} + \tilde{\lambda}(\psi) \, 
Y_{\alpha\beta} \tilde{Y}^{\alpha\beta}\biggr]
\label{act1}
\end{equation}
where  both, $\lambda(\psi)$ and $\tilde{\lambda}(\psi)$ depend explicitly on the field $\psi$ whose action and 
potential can be specified in a given scenario; furthermore, in Eq. (\ref{act1}) 
\begin{equation}
Y_{\alpha\beta} = \nabla_{[\alpha}Y_{\beta]} = \partial_{[\alpha} Y_{\beta]},\qquad 
\tilde{Y}^{\alpha\beta} = \frac{\epsilon^{\alpha\beta\rho\sigma}}{2 \sqrt{-G}} Y_{\rho\sigma}.
\label{act1a} 
\end{equation}
The explicit form of the field strengths in terms of the electric and magnetic fields
are given, respectively, by $Y_{0i} = a^2\, e_{i}$ and by 
$Y_{ij} = - a^2 \, \epsilon_{ijk} \,b^{k}$ while the dynamical equations and the Bianchi identities 
are 
\begin{equation}
\nabla_{\alpha} \biggl(\lambda Y^{\alpha\beta} + \tilde{\lambda} \tilde{Y}^{\alpha\beta}\biggr) =0,\qquad 
\nabla_{\alpha} \tilde{Y}^{\alpha\beta} =0.
\label{act1b}
\end{equation}
 In terms of the canonical 
hyperelectric and hypermagnetic fields the action becomes:
\begin{equation}
S_{Y} = \frac{1}{2} \int d^{4} x [ \vec{E}^2 - \vec{B}^2],\qquad \vec{B} = \sqrt{\frac{\lambda}{4\pi}} \, a^2 \, \vec{b}, \qquad 
\vec{E} = \sqrt{\frac{\lambda}{4\pi}} \, a^2 \, \vec{e};
\label{act2}
\end{equation}
the coupling $\tilde{\lambda}(\psi)$ will now be set to zero even if, as we shall remark, the 
considerations on the duality of the electromagnetic spectra obtained from the parametric amplification of vacuum fluctuations 
can be generalized to the case where $\tilde{\lambda} \neq 0$.
In the absence of charged particles the evolution equations of $\vec{E}$ and $\vec{B}$ can be written as
\begin{equation}
\vec{\nabla} \times( \sqrt{\lambda} \vec{B}) - \frac{\partial}{\partial \tau} ( \sqrt{\lambda} \vec{E}) =0, \qquad 
\frac{\partial}{\partial \tau}\biggl( \frac{\vec{B}}{\sqrt{\lambda}}\biggr) + \vec{\nabla} 
\times \biggl(\frac{\vec{E}}{\sqrt{\lambda}}\biggr) =0.
\label{dual1}
\end{equation}
Under electric-magnetic duality, i.e. 
 \begin{equation}
 \vec{E} \to - \vec{B}, \qquad \vec{B} \to \vec{E}, \qquad \sqrt{\lambda} \to 
 \frac{1}{\sqrt{\lambda}},
 \label{dual3}
 \end{equation}
the first of Eq. (\ref{dual1}) goes into the second (and, vice-versa, the second equation is transformed into the 
first one). In this sense $\sqrt{\lambda}$ plays here the role of inverse coupling
$g_{Y}^2/(4\pi) = \alpha_{Y} = 1/\sqrt{\lambda}$. Note that the action (\ref{dual1}) is not invariant 
under (\ref{dual3}) (see also \cite{DT} for detailed discussions on the canonical formalism applied to duality). 
In the  present context the transformation of Eq. (\ref{dual3}) will simply be used to compute the electric and magnetic power spectra when the gauge coupling evolves with the conformal time. 
If $\tilde{\lambda} \neq 0 $ in Eq. (\ref{act1}) the resulting theory can be shown 
to possess a continuous $SL(2,R)$ symmetry which generalizes  (\ref{dual3}). Some of the considerations illustrated here can be generalized to the full $SL(2,R)$ case but this will not be our primary purpose.

The plan of the present paper is the following. In section \ref{sec2} it will be shown how the electric 
and magnetic power spectra can be related by duality transformations both exactly and approximately 
(i.e. when the relevant wavelengths are larger than the Hubble radius). The analysis 
will be conducted primarily in the case of single field inflationary models even if occasional remarks on different models 
will be made. In section \ref{sec3} it will be shown how the finite value of the conductivity breaks 
the duality symmetry and distinguishes  the magnetic degrees of freedom from the electric ones. In a more formal 
sense the presence of the conductivity induces an effective decoherence of the amplified electromagnetic fluctuations. 
In section \ref{sec4} numerical examples will be discussed and the parameter space of the model will be 
be studied. The concluding remarks are collected in section \ref{sec5}.  

\renewcommand{\theequation}{2.\arabic{equation}}
\setcounter{equation}{0}
\section{Electric-magnetic duality and power spectra}
\label{sec2}
In time-dependent (and conformally flat) backgrounds such as the one of Eq. (\ref{met}) the quantization 
 can be easily performed, for the present ends, in the Coulomb gauge. The Coulomb gauge condition (i.e. $Y_{0} =0$  and $\vec{\nabla}\cdot \vec{Y} =0$) is preserved under 
a conformal rescaling of the metric (\ref{met}); the Lorentz 
gauge condition does not have the same property \cite{LF}. Furthermore, 
in the Coulomb gauge, the duality properties are easily discussed \footnote{It is of course possible to keep 
also $Y_{0}$ and the longitudinal part of the Abelian vector potential, i.e. 
$\vec{Y}_{\mathrm{L}}$. It can be easily shown, however, that the transverse 
variables are decoupled from the longitudinal and from the gauge contributions. Indeed, without any specific gauge fixing, the extremization of the action with respect to $Y_{0}$ implies that $\vec{Y}^{\,\,'}_{\mathrm{L}} = \vec{\nabla} Y_{0}$.}
\cite{DT}. The action of Eq. (\ref{act1})  can then be written as
\begin{equation}
S_{Y} = \int \,d\tau\, L_{Y}(\tau), \qquad L_{Y}(\tau) = \int d^{3} x\, {\mathcal L}_{Y}(\vec{x},\tau),
\label{dual5}
\end{equation}
where, in the case $\tilde{\lambda} =0$,  
\begin{equation}
{\mathcal L}_{Y}(\vec{x},\tau) = \frac{1}{2} \biggl\{ \vec{y}^{\,\prime \,2} + \biggl[\frac{(\sqrt{\lambda})'}{\sqrt{\lambda}}\biggr]^2 
 \vec{y}^{\,2}  - 2 \frac{(\sqrt{\lambda})'}{\sqrt{\lambda}} \vec{y} \cdot \vec{y}^{\,\prime} - \partial_{i} \vec{y} \cdot \partial^{i} \vec{y}\biggr\},
\label{dual6}
\end{equation}
and $\vec{y} = \sqrt{\lambda/(4\pi)} \vec{Y}$. The canonical momentum conjugate to $\vec{y}$ can be easily 
obtained from Eq. (\ref{dual6}) and it coincides, up to a sign, with the canonical electric field, i.e. 
\begin{equation}
\vec{\pi} = \vec{y}^{\,\prime} - \frac{(\sqrt{\lambda})'}{\sqrt{\lambda}} \vec{y} = - \vec{E} 
\label{dual7}
\end{equation}
while the relation of $\vec{y}$ to $\vec{B}$ is simply given by $\vec{B} = \vec{\nabla}\times \vec{y}$. The canonical 
Hamiltonian is then given by 
\begin{equation}
H_{Y}(\tau) = \frac{1}{2} \int d^3 x \biggl[ \vec{\pi}^{2} + 2 \frac{(\sqrt{\lambda})'}{\sqrt{\lambda}} \vec{\pi} \cdot \vec{y} + 
\partial_{i} \vec{y} \cdot \partial^{i} \vec{y}\biggr].
\label{dual8}
\end{equation}
The Fourier mode expansion for the canonical fields
\begin{equation}
\vec{\pi}(\vec{x},\tau) = \frac{1}{(2\pi)^{3/2}} \int d^{3} k\,\, \vec{\pi}_{\vec{k}}(\tau) \,\,e^{-i \vec{k}\cdot\vec{x}}, \qquad 
 \vec{y}(\vec{x},\tau) = \frac{1}{(2\pi)^{3/2}} \int d^{3} k \,\, \vec{y}_{\vec{k}}(\tau) \,\,e^{-i \vec{k}\cdot\vec{x}},
 \label{dual9}
 \end{equation}
can be inserted into Eq. (\ref{dual8}) and the resulting expression is
\begin{equation}
H_{Y}(\tau) = \frac{1}{2} \int d^3 k \biggl[ \vec{\pi}_{\vec{k}} \cdot \vec{\pi}_{-\vec{k}} +  \frac{(\sqrt{\lambda})'}{\sqrt{\lambda}}
\biggl( \vec{\pi}_{\vec{k}} \cdot \vec{y}_{-\vec{k}}  +  \vec{\pi}_{-\vec{k}} \cdot \vec{y}_{\vec{k}}\biggr)
+k^2 \vec{y}_{\vec{k}} \cdot \vec{y}_{-\vec{k}}\biggr].
\label{dual10}
\end{equation}
Equation (\ref{dual10}) is invariant under the transformation 
\begin{eqnarray}
&& \vec{\pi}_{\vec{k}} \to - k \vec{y}_{\vec{k}}, \qquad \vec{y}_{\vec{k}} \to \frac{1}{k} \vec{\pi}_{\vec{k}},\qquad 
\sqrt{\lambda} \to \frac{1}{\sqrt{\lambda}},
\nonumber\\
&&  \vec{\pi}_{-\vec{k}} \to - k \vec{y}_{-\vec{k}}, \qquad \vec{y}_{-\vec{k}} \to \frac{1}{k} \vec{\pi}_{-\vec{k}},
\label{dual11}
\end{eqnarray}
where $k = |\vec{k}|$. The Hamilton equations derived from Eq. (\ref{dual10}) become:
\begin{eqnarray}
&&\vec{y}_{\vec{k}}^{\,\prime} = \vec{\pi}_{\vec{k}} + \frac{(\sqrt{\lambda})'}{\sqrt{\lambda}} \vec{y}_{\vec{k}},
\label{dual12}\\ 
&&\vec{\pi}_{\vec{k}}^{\,\prime} = - k^2 \vec{y}_{\vec{k}} -\frac{(\sqrt{\lambda})'}{\sqrt{\lambda}} \vec{\pi}_{\vec{k}}.
\label{dual13}
\end{eqnarray}
By applying the transformation (\ref{dual11}), Eq. (\ref{dual12}) is transformed into Eq. (\ref{dual13}) and vice versa.

In cosmological backgrounds, any canonical transformation depending 
explicitly upon time does modify the form of the Hamiltonian without altering the Hamilton equations. 
While such an occurrence is harmless at the level of the evolution equations, 
the different forms of the Hamiltonian can make the difference when setting initial conditions 
at a finite time during the inflationary epoch. A similar problem arises  for spin-0 \cite{mg3} and spin-2 fluctuations 
of the geometry \cite{mg4} and it is related to the so-called transplankian 
ambiguity of inflationary initial conditions. In the present context pre-inflationary 
initial conditions can induce explicit sources which break the symmetry  of the system of equations 
given in Eqs. (\ref{dual12}) and (\ref{dual13}). 

The canonical fields can be promoted to quantum operators (i.e. $y_{i} \to \hat{y}_{i}$ and $\pi_{i} \to \hat{\pi}_{i}$) 
obeying (equal time) commutation relations\footnote{Units $\hbar=c= \kappa_{\mathrm{B}} = 1$ will be assumed hereunder.}:
\begin{equation}
[\hat{y}_{i}(\vec{x}_{1},\tau),\hat{\pi}_{j}(\vec{x}_{2},\tau)] = i \Delta_{ij}(\vec{x}_{1} - \vec{x}_{2}),\qquad 
\Delta_{ij}(\vec{x}_{1} - \vec{x}_{2}) = \int \frac{d^{3}k}{(2\pi)^3} e^{i \vec{k} \cdot (\vec{x}_{1} - \vec{x}_2)} P_{ij}(k), 
\label{dual17}
\end{equation}
where $P_{ij}(k) = (\delta_{ij} - k_{i} k_{j}/k^2)$. The function $\Delta_{ij}(\vec{x}_{1} - \vec{x}_{2})$ is the transverse generalization of the Dirac delta function and such an extension is 
mandatory since $ \vec{\nabla} \cdot \vec{E} =0$ because of the Gauss constraint and 
$\vec{\nabla}\cdot \vec{y}=0$ because of the gauge condition\footnote{Notice that $\lambda$ and $\tilde{\lambda}$ are 
here assumed to depend only upon the conformal time coordinate $\tau$.}. The field operators will then become
\begin{eqnarray}
\hat{y}_{i}(\vec{x},\tau) = \int\frac{d^{3} k}{(2\pi)^{3/2}} \sum_{\alpha} e^{(\alpha)}_{i}(k) \, 
\biggl[ f_{k}(\tau) \, \hat{a}_{k,\alpha} e^{- i \vec{k} \cdot\vec{x}} +  f_{k}^{*}(\tau) \, \hat{a}^{\dagger}_{k,\alpha} e^{ i \vec{k} \cdot\vec{x}}\biggr],
\label{dual18}\\
\hat{\pi}_{i}(\vec{x},\tau) = \int\frac{d^{3} k}{(2\pi)^{3/2}} \sum_{\alpha} e^{(\alpha)}_{i}(k) \, 
\biggl[ g_{k}(\tau) \, \hat{a}_{k,\alpha} e^{- i \vec{k} \cdot\vec{x}} +  g_{k}^{*}(\tau) \, \hat{a}^{\dagger}_{k,\alpha} e^{ i \vec{k} \cdot\vec{x}}\biggr],
\label{dual19}
\end{eqnarray}
where, from Eq. (\ref{dual13}), $f_{k}$ and $g_{k}$ obey:
\begin{equation}
f_{k}' = g_{k} + \frac{(\sqrt{\lambda})'}{\sqrt{\lambda}} f_{k},\qquad 
g_{k}' = - k^2 f_{k} - \frac{(\sqrt{\lambda})'}{\sqrt{\lambda}} g_{k}.
\label{dual21}
\end{equation}
The mode functions $f_{k}(\tau)$ and $g_{k}(\tau)$ must also satisfy (Wronskian) normalization condition which follows from enforcing the canonical commutators:
\begin{equation}
f_{k}(\tau)g_{k}^{*}(\tau) - f_{k}^{*}(\tau) g_{k}(\tau) =i.
\label{dual22}
\end{equation}
In Eqs. (\ref{dual18}) and (\ref{dual19}) $e^{(\alpha)}_{i}(k)$ must obey 
$\sum_{\alpha} \, e^{(\alpha)}_{i}(k) \,e^{(\alpha)}_{j}(k) = P_{ij}(k)$. Finally 
the creation and annihilation operators appearing in Eqs. (\ref{dual18}) 
and (\ref{dual19}) obey $[\hat{a}_{\vec{k},\alpha}, \hat{a}^{\dagger}_{\vec{p}\,\beta} ] = \delta_{\alpha\beta} \delta^{(3)}(\vec{k} + \vec{p})$.
The state annihilated by $\hat{a}_{k,\alpha}$, i.e. $|0\rangle$, 
is indeed the vacuum but only in the limit $\tau\to -\infty$. Such a limit is, strictly speaking,  not physical since expanding de Sitter space is not geodesically complete. This consideration implies that the normalization procedure must 
be conducted more carefully whenever the number of inflationary e-folds is close to minimal. Corrections 
depending upon the finite normalization time can appear in the power spectrum \cite{mg3,mg4}. These 
corrections will be neglected here but can be important in more refined treatments. 
The two point  function of the electric and magnetic fields for equal times but 
different spatial locations is \cite{variation1}:
\begin{eqnarray}
&& \langle 0| \hat{B}_{i}(\vec{x},\tau) \, \hat{B}_{j}( \vec{x} + \vec{r},\tau)|0 \rangle = \int d\ln{k} {\mathcal P}_{\mathrm{B}}(k,\tau) 
\, P_{ij}(k) \frac{\sin{k r}}{kr},
\label{dual15}\\
&& \langle 0| \hat{E}_{i}(\vec{x},\tau) \, \hat{E}_{j}( \vec{x} + \vec{r},\tau)|0 \rangle = \int d\ln{k} {\mathcal P}_{\mathrm{E}}(k,\tau) 
\, P_{ij}(k) \frac{\sin{k r}}{kr},
\label{dual16}
\end{eqnarray}
where ${\mathcal P}_{\mathrm{B}}(k,\tau)$ and ${\mathcal P}_{\mathrm{E}}(k,\tau)$ are, by definition, the magnetic and electric power spectra:
\begin{equation}
 {\mathcal P}_{\mathrm{E}}(k,\tau) = \frac{k^5}{2\pi^2} |f_{k}(\tau)|^2,
\qquad 
{\mathcal P}_{\mathrm{B}}(k,\tau) = \frac{k^3}{2\pi^2} |g_{k}(\tau)|^2.
\label{dual24}
\end{equation}
The expectation values of the energy density as well as of all the 
other components of the energy-momentum tensor can be derived in similar ways. The energy momentum tensor can be  written directly in terms of the canonical fields $\vec{E}$ and $\vec{B}$ introduced in Eq. (\ref{act2}):
\begin{eqnarray}
&& T_{0}^{0} = \frac{1}{2 a^4} \biggl( \vec{E}^2 + \vec{B}^2 \biggr), \qquad T_{0}^{i} = \frac{1}{a^4} (\vec{E} \times \vec{B})^{i}
\label{dual25}\\
&& T_{i}^{j} = - \frac{1}{6 a^4}  \biggl( \vec{E}^2 + \vec{B}^2 \biggr)\delta_{i}^{j} + \frac{1}{a^4} \biggl( E_{i} E^{j} - \frac{\vec{E}^2}{3}\delta_{i}^{j}\biggr) + \frac{1}{a^4} \biggl( B_{i} B^{j} - \frac{\vec{B}^2}{3}\delta_{i}^{j}\biggr).
\label{dual26}
\end{eqnarray}
By averaging the different components of the energy-momentum tensor we get:
and it is given by:
\begin{eqnarray}
&& \langle 0| T_{0}^{0} |0\rangle =\rho_{Y}(\tau) =  \frac{1}{a^4} \int d\ln{k} \biggl[{\mathcal P}_{\mathrm{E}}(k,\tau) +
 {\mathcal P}_{\mathrm{B}}(k,\tau)\biggr],
\nonumber\\ 
&& \langle 0| T_{i}^{j} |0\rangle = - p_{Y}(\tau) \delta_{i}^{j}, \qquad p_{Y} = \frac{\rho_{Y}}{3}, 
\label{dual27}
\end{eqnarray} 
where it has been used that, for a generic function $h(k,\tau)$, 
\begin{equation}
\int d^{3}k\,\, k_{i} k^{j} h(k, \tau) = \frac{4 \pi}{3} \delta_{i}^{j} \int k^{5}\, h(k,\tau)\,  d\ln{k},\qquad \int d^{3}k k^{i} h(k,\tau) =0.
\nonumber
\end{equation}
When the gauge coupling evolves in time as
 \begin{equation}
\alpha_{Y}(\tau) = \alpha_{Y}(\tau_{i})\biggl(-\frac{\tau}{\tau_{\mathrm{i}}}\biggr)^{2\nu - 1}, \qquad g_{Y}(\tau) = 
g_{Y}(\tau_{i}) \biggl(-\frac{\tau}{\tau_{\mathrm{i}}}\biggr)^{\nu - 1/2}, 
\label{dual28}
\end{equation}
the solution of Eq. (\ref{dual21}) can be written  as  
\begin{eqnarray}
&& f_{k}(\tau) = \frac{{\mathcal N}}{\sqrt{2 k}} \, \sqrt{- k \tau} \, H_{\nu}^{(1)}(- k \tau),\qquad {\mathcal N} = \sqrt{\frac{\pi}{2}} e^{i \pi( \nu +1/2)/2},
\label{dual32}\\
&& g_{k}(\tau) = - {\mathcal N}\, \sqrt{\frac{k}{2}}\, \sqrt{-k\tau} \, H_{\nu-1}^{(1)}(- k \tau).
\label{dual33}
\end{eqnarray}
Equations (\ref{dual32}) and (\ref{dual33}) obey the Wronskian normalization condition 
of Eq. (\ref{dual22}) as it can be easily verified by recalling the properties of the Hankel functions \cite{abr1,abr2}. In the limit $\tau \to - \infty$ Eqs. (\ref{dual32})--(\ref{dual33}) become 
\begin{equation}
\lim_{\tau \to - \infty} \, f_{k}(\tau) = \frac{1}{\sqrt{2 k}} e^{- i k \tau},\qquad \lim_{\tau \to - \infty} g_{k}(\tau) = 
- i \sqrt{\frac{k}{2}} e^{- i k \tau}.
\label{dual35}
\end{equation}
If $\nu = 1/2$, $\lambda$ is constant and the asymptotic solutions of Eq. (\ref{dual35}) are actually  
exact solutions. When $\nu= 0$ $f_{k}(\tau)$ diverges logarithmically in the limit of small argument, i.e. 
$|k\tau|\ll 1$. 
Using Eqs. (\ref{dual32}) and (\ref{dual33}) the electric and magnetic power spectra of Eq. (\ref{dual24}) become
\begin{eqnarray}
{\mathcal P}_{\mathrm{E}}(k,\tau) = \frac{k^4}{8\pi} (- k\tau) |H_{\nu-1}^{(1)}(-k\tau)|^2,\qquad 
{\mathcal P}_{\mathrm{B}}(k,\tau) = \frac{k^4}{8\pi} (- k\tau) |H_{\nu}^{(1)}(-k\tau)|^2.
\label{dual36}
\end{eqnarray}
Let us now perform a duality transformation and show that the electric and the magnetic power 
spectra are interchanged while the energy density is invariant. Consider, therefore, the following transformation 
\begin{equation}
g_{Y}(\tau) \to \tilde{g}_{Y}(\tau) = \frac{1}{g_{Y}(\tau)}, \qquad \qquad \tilde{g}_{Y}(\tau) = 
\tilde{g}_{Y}(\tau_{i}) \biggl(-\frac{\tau}{\tau_{\mathrm{i}}}\biggr)^{\tilde{\nu} - 1/2}.
\label{dual37}
\end{equation}
By repeating the same steps discussed above with the transformed gauge coupling it is 
clear that 
\begin{equation}
\tilde{{\mathcal P}}_{\mathrm{E}}(k,\tau) = \frac{k^4}{8\pi} (- k\tau) |H_{\tilde{\nu}-1}^{(1)}(-k\tau)|^2,\qquad 
\tilde{{\mathcal P}}_{\mathrm{B}}(k,\tau) = \frac{k^4}{8\pi} (- k\tau) |H_{\tilde{\nu}}^{(1)}(-k\tau)|^2.
\label{dual38}
\end{equation}
But Eqs. (\ref{dual28}) and (\ref{dual38}) imply that 
\begin{equation}
\tilde{\nu} = 1 - \nu, \qquad \tilde{{\mathcal P}}_{\mathrm{E}}(k,\tau) = {\mathcal P}_{\mathrm{B}}(k,\tau), \qquad 
\tilde{{\mathcal P}}_{\mathrm{B}}(k,\tau) = {\mathcal P}_{\mathrm{E}}(k,\tau).
\label{dual39}
\end{equation}
The result of Eqs. (\ref{dual38}) and (\ref{dual39}) follow immediately by appreciating that \cite{abr1,abr2} 
\begin{equation}
|H_{\nu-1}^{(1)}(-k\tau)|^2 =  |H_{1-\nu}^{(1)}(-k\tau)|^2, \qquad H_{-\mu}^{(1)}(z) = e^{i \mu\pi}H_{\mu}^{(1)}(z)
\label{dual40}
\end{equation}
where $\mu$ and $z$ denote, respectively, a generic (real) index and a generic argument of the Hankel function. 
The second relation of Eq. (\ref{dual40}) simply states that a generic Hankel function of first kind 
with index $-\mu$ equals the corresponding Hankel function of index $+\mu$ up to a $\mu$-dependent phase 
which is immaterial  as far as the square moduli are concerned. For the Hankel function of second kind the same kind of relation holds
and it follows by taking the complex conjugate of the second relation in Eq. (\ref{dual40}). 
On the basis of the obtained results it is also clear that under Eq. (\ref{dual37}) 
\begin{eqnarray}
&& \rho_{Y}(\tau) \to \tilde{\rho}_{Y}(\tau) = \frac{1}{a^4}\int d\ln{k}  \biggl[\tilde{{\mathcal P}}_{\mathrm{E}}(k,\tau) +
 \tilde{{\mathcal P}}_{\mathrm{B}}(k,\tau)\biggr] = \rho_{Y}(\tau),
\nonumber\\ 
&& p_{Y} \to \tilde{p}_{Y}(\tau) = p_{Y}(\tau).
\label{dual41}
\end{eqnarray}
It is useful to define, for immediate peruse, the spectral energy density in units of the critical energy density
\begin{equation}
\Omega_{Y}(k,\tau) = \frac{1}{\rho_{\mathrm{crit}}} \frac{d \rho_{Y}}{d\ln{k}} = \frac{8\pi}{3\, a^4\, H^2 M_{\mathrm{P}}^2} \biggl[ P_{\mathrm{E}}(k,\tau) + P_{\mathrm{B}}(k,\tau)\biggr],
\label{dual42}
\end{equation}
where $\rho_{\mathrm{crit}} = 3 H^2/(8\pi G)$ is the critical energy density. Using the results of Eq. (\ref{dual41}),
$\tilde{\Omega}_{Y}(k,\tau) = \Omega_{Y}(k,\tau)$. 

Consider now the case when the evolution of the geometry evolves according to a quasi-de Sitter stage of expansion driven by a (single) scalar degree of freedom $\varphi$. After the quasi-de Sitter stage the conventional lore 
suggests that the radiation dominated stage of expansion takes place almost suddenly. In what follows 
the latter possibility will be examined together with different physical scenarios like the ones 
stipulating that the transition to radiation has a finite duration (see also section \ref{sec3}). 
It will be also worth examining, for the 
present ends, all the cases when the transition to radiation is delayed by a long phase where 
the speed of sound of the plasma is either softer of stiffer than radiation (see also section \ref{sec4}). 

During the quasi-de Sitter stage of expansion the relevant slow-roll parameters are given by
\begin{equation}
\epsilon = - \frac{\dot{H}}{H^2} \ll 1, \qquad \eta = \frac{\ddot{\varphi}}{H\dot{\varphi}}\ll 1.
\label{dual29}
\end{equation}
Denoting with  $V(\varphi)$  the inflaton potential, the slow-roll equations can be written in the form 
\begin{equation}
H^2 M_{\mathrm{P}}^2  \simeq \frac{8\pi}{3} V, \qquad 3 H \dot{\varphi} + \frac{\partial V}{\partial \varphi} \simeq 0.
\label{dual30}
\end{equation}
Using Eqs. (\ref{dual29}) and (\ref{dual30}) it is rather standard to derive the following relations 
\begin{equation}
\epsilon =  \frac{\overline{M}^2_{\mathrm{P}}}{2} \biggl(\frac{V_{,\varphi}}{V}\biggr)^2, \qquad \eta = \epsilon - \overline{\eta},\qquad 
\overline{\eta} = \overline{M}_{\mathrm{P}}^2 \frac{V_{,\varphi\varphi}}{V}.
\label{SR1}
\end{equation}
where the reduced Planck mass $\overline{M}_{\mathrm{P}} = M_{\mathrm{P}}/\sqrt{8\pi}$ has been introduced. 
To make the forthcoming statements quantitative (rather than, as sometimes done, just 
qualitative) we recall that the power spectra of scalar and tensor fluctuations are assigned as 
\begin{equation}
{\mathcal P}_{{\mathcal R}}(k) = {\mathcal A}_{\mathcal R} \biggl(\frac{k}{k_{\mathrm{p}}}\biggr)^{n_{\mathrm{s}} -1}, \qquad 
{\mathcal P}_{\mathrm{T}} = r_{\mathrm{T}} {\mathcal A}_{{\mathcal R}}
\biggl(\frac{k}{k_{\mathrm{p}}}\biggr)^{n_{\mathrm{T}}},
\label{scten}
\end{equation}
where ${\mathcal A}_{{\mathcal R}}$ and $r_{\mathrm{T}} {\mathcal A}_{{\mathcal R}}$ are, respectively, the 
amplitudes of the scalar and of the tensor modes at the pivot scale $k_{\mathrm{p}} =0.002\, \mathrm{Mpc}^{-1}$
and where the spectral indices are related to $\epsilon$ and $\overline{\eta}$ as well as to $r_{\mathrm{T}}$ 
\begin{equation}
n_{\mathrm{T}} = - 2\epsilon,\qquad n_{\mathrm{s}} = 1 - 6\epsilon + 2 \overline{\eta},\qquad r_{\mathrm{T}} = 16 \epsilon = - 8 n_{\mathrm{T}}.
\end{equation}
The WMAP 5yr data 
alone in the light of the vanilla $\Lambda$CDM scenario imply 
\cite{WMAP51} ${\mathcal A}_{{\mathcal R}} = (2.41 \pm 0.11)\times 10^{-9}$ and provide an upper limit on $r_{\mathrm{T}}$ and $\epsilon$, i.e. 
$r_{\mathrm{T}} < 0.58$ and $\epsilon< 0.036$ (95 \% CL). The bounds on $\epsilon$ can be improved by combining the WMAP5yr data with other data sets. For the present purposes, explicit 
numerical estimates will assume, as it will be seen, ${\mathcal A}_{{\mathcal R}} = 2.41\times 10^{-9}$ and $\epsilon = 0.02$ implying 
\begin{equation}
\frac{H^2}{M_{\mathrm{P}}^2} = 1.51\times 10^{-10} \biggl(\frac{\epsilon}{0.02}\biggr) \biggl(\frac{{\mathcal A}_{{\mathcal R}}}{2.41\times 10^{-9}}\biggr).
\label{dual50}
\end{equation}
The slow-roll parameters enter the relation between the conformal time 
coordinate $\tau$ and the Hubble rate. More precisely, recall that
\begin{equation}
\tau = \int \frac{dt}{a(t)} = - \frac{1}{a H} + \epsilon \int \frac{d a}{a^2 H}
\label{tauHa}
\end{equation}
where the second equality follows after integration by parts assuming that $\epsilon$ is 
constant (as it happens in the case when the potential, at least locally, can be 
approximated with a monomial in $\varphi$). Since $\int dt/a = \int da/(a^2 H)$, Eq. (\ref{tauHa}) allows to express $a H$ in terms of $\tau$ and $\epsilon$, i.e. $a H = - 1/[\tau ( 1 - \epsilon)]$. Using the latter 
relation Eq. (\ref{dual42}) can be expressed as 
\begin{equation}
\Omega_{Y}(k,\tau,\nu) = \frac{H^2}{3 M_{\mathrm{P}}^2} \frac{x^5}{(1 - \epsilon)}\biggl[ \bigg|H^{(1)}_{\nu}\biggl(\frac{x}{1 - \epsilon}\biggr)\bigg|^2 + \bigg|H^{(1)}_{\nu -1}\biggl(\frac{x}{1 -\epsilon}\biggr)\bigg|^2\biggr],
\label{dual43}
\end{equation}
where $\epsilon$ is the slow-roll parameter introduced in Eq. (\ref{dual29}) and the notation 
\begin{equation}
-k\tau = \frac{k}{aH (1-\epsilon)} = \frac{x}{1 - \epsilon}.
\label{dual44}
\end{equation}
will now be employed. An interesting limiting case is the one where $x/(1 -\epsilon)\ll 1$ where the explicit expression of $\Omega_{Y}(x,\nu)$ 
can be evaluated by using the small argument limit of the Hankel functions. 
This limit corresponds to physical wavelengths which are 
larger than the Hubble radius during the quasi-de Sitter stage of expansion 
and in the absence of Ohmic currents.
For practical purposes it is useful to define the function 
\begin{equation}
{\mathcal C}(z, \epsilon) = \frac{\Gamma^2(z)}{3 \pi^2} (1 - \epsilon)^{2 z -1} 2^{2 z} 
\label{def}
\end{equation}
where $z$ is a real number which is otherwise unrestricted.
If $\nu > 1/2 $ the spectral energy density will be given, respectively, by 
\begin{eqnarray}
\Omega^{(\mathrm{a})}_{Y}(x,\nu) &=& \frac{H^2}{M_{\mathrm{P}}^2} {\mathcal C}(\nu,\epsilon) x^{5 - 2 \nu} \, 
 \biggl[1 + \frac{\Gamma^2(\nu-1)\, x^{2}}{4 \Gamma^2(\nu) ( 1 - \epsilon)^2}\biggr],\qquad \nu > 1,
\label{dual45}\\
\Omega^{(\mathrm{b})}_{Y}(x,\nu) &=& \frac{H^2}{M_{\mathrm{P}}^2} {\mathcal C}(\nu,\epsilon)  \, 
x^{5 - 2 \nu} \biggl[1 + \frac{\Gamma^2(1 - \nu)}{\Gamma^2(\nu)}\frac{2^{2 - 4 \nu}}{(1- \epsilon)^{4\nu -1}}\, x^{4 \nu -2}\biggr],\qquad \frac{1}{2}< \nu <1.
\label{dual46}
\end{eqnarray}
The case of increasing gauge coupling can be immediately obtained from Eqs. (\ref{dual45}) and (\ref{dual46}). Indeed, recalling 
the transformation of Eq. (\ref{dual37})  and bearing in mind the relation $\tilde{\nu} = 1 - \nu$ we shall have that 
\begin{eqnarray}
\Omega^{(\mathrm{c})}_{Y}(x,\nu) &=&  \frac{H^2}{M_{\mathrm{P}}^2} {\mathcal C}(1-\nu, \epsilon)
x^{3 + 2\nu}\biggl[ 1 + \frac{\Gamma^2(-\nu)\,x^2}{4\Gamma^2(1-\nu)(1-\epsilon)^2}\biggr],\qquad \nu<0,
\label{dual47}\\
\Omega^{(\mathrm{d})}_{Y}(x,\nu) &=& \frac{H^2}{M_{\mathrm{P}}^2}  {\mathcal C}(1-\nu, \epsilon) x^{3 + 2 \nu} \biggl[ 1+ 
\frac{\Gamma^2(\nu) 2^{4\nu -2}}{\Gamma^2(1-\nu)} \frac{x^{2 - 4\nu}}{(1-\epsilon)^{2 - 4 \nu}}\, \biggr], \qquad 0< \nu < \frac{1}{2},
\label{dual48} 
\end{eqnarray}
where the cases $\nu < 0$ and $0<\nu <1/2$ correspond, respectively, to $\tilde{\nu} >1$ and to $1/2 < \tilde{\nu} < 1$. 
It is useful to notice that the duality of the spectra can be productively used 
to relate the four cases discussed above which reduce effectively to two cases 
since, as it can be easily seen, 
\begin{equation}
\Omega^{(\mathrm{c})}_{Y}(x,\nu) =   \Omega^{(\mathrm{a})}_{Y}(x,1-\nu),\qquad \Omega^{(\mathrm{d})}_{Y}(x,\nu) =   \Omega^{(\mathrm{b})}_{Y}(x,1-\nu).
\label{dualrel}
\end{equation}
The transformation $\nu \to \tilde{\nu} = 1 - \nu$ implies that in the cases 
$\tilde{\nu}=0$ and $\nu =0$ the spectral energy density is the same  
and it is given by:
\begin{equation}
\Omega_{Y}(x,\nu) =  \frac{H^2}{M_{\mathrm{P}}^2} x^3 {\mathcal C}(1, \epsilon) \biggl[ 1 + \frac{x^2}{(1-\epsilon)^2} \ln^2{\biggl(\frac{x}{1-\epsilon}\biggr)}\biggr].
\label{dual49}
\end{equation}
Concerning  Eqs. (\ref{dual45})--(\ref{dual46}) and Eqs.  (\ref{dual47})--(\ref{dual48}) few comments are in order.
It can be easily verified that the spectral energy density is increasing for $- 3/2< \nu < 5/2$ while it is decreasing for $\nu < -3/2$ and for $\nu > 5/2$. The case of flat spectrum can be realized both for $\nu = -3/2$ as well as for $\nu = 5/2$;
finally the closure bound seems to imply, grossly speaking, that
\begin{equation}
\nu > -\frac{3}{2} - \omega, \qquad \nu < \frac{5}{2} + \omega, \qquad \omega \simeq 0.3.
\label{clnaive}
\end{equation}
The estimate of Eq. (\ref{clnaive}) follows by recalling that
when the spectra are increasing the most relevant constraint stems from $x_{\mathrm{max}} \simeq 1$; 
when the spectra are decreasing the critical density bound should be applied 
for $x= x_{\mathrm{min}}$ where 
\begin{equation}
x_{\mathrm{min}} = 3.149\times 10^{-30} \biggl(\frac{H}{M_{\mathrm{P}}}\biggr)^{\mu -1} 
\biggl(\frac{H_{\mathrm{r}}}{M_{\mathrm{P}}}\biggr)^{1/2 -\mu}.
\label{xmin1}
\end{equation}
The estimate of Eq. (\ref{xmin1}) takes into account an intermediate (expanding) phase 
characterized by $\mu = 2/[3(w_{\mathrm{t}} +1)]$ where $w_{\mathrm{t}}$ is the barotropic 
index characterizing the intermediate phase while $H_{\mathrm{r}}$ marks the onset of the radiation-dominated 
stage of expansion. In the limit of sudden reheating  $H_{\mathrm{r}} = H$.
By neglecting all the numerical pre-factors in Eqs. (\ref{dual45})--(\ref{dual46}) and Eqs.  (\ref{dual47})--(\ref{dual48})
the value of $\omega$ can be estimated as:
\begin{equation}
\omega= \frac{\log{(H/M_{\mathrm{P}})}}{\log{x_{\mathrm{min}}}}.
\label{xmin2}
\end{equation}
By taking into account all the pre-factors the estimate $\omega \sim 0.23$ for $\epsilon =0.02$. Note that 
the estimate of Eq. (\ref{xmin2}) is based, just for illustration, on the sudden reheating approximation, i.e. 
$H_{\mathrm{r}} = H$ in Eq. (\ref{xmin1}). The limits stemming from Eqs. (\ref{xmin2}) 
(as well as from Eq. (\ref{clnaive})) just mean that, as soon as $\nu$ gets larger than $-(3/2 +\omega)$ (or, equivalently, 
smaller than $5/2+ \omega$) the dynamics of the gauge coupling should be discussed self-consistently  by using the 
following set of equations
\begin{eqnarray}
&&\overline{M}_{\mathrm{P}}^2 {\mathcal H}^2 = \frac{1}{3}\biggl[\frac{{\varphi'}^2}{2} + V a^2\biggr] 
+ \frac{1}{3}\biggl[\frac{{\psi'}^2}{2} + W a^2\biggr]  + \rho_{Y},
\nonumber\\
&& \overline{M}_{\mathrm{P}}^2 ({\mathcal H}^2 - {\mathcal H}') = \frac{1}{2} ({\varphi'}^2 + {\psi'}^2) + \frac{2}{3} \rho_{Y},
\nonumber\\
&& \varphi'' + 2 {\mathcal H} \varphi' + \frac{\partial V}{\partial\varphi} a^2 =0,
\nonumber\\
&& \psi'' + 2  {\mathcal H} \psi'  +  \frac{\partial W}{\partial\psi} a^2 + \frac{1}{2 a^2} \frac{\partial \ln{\lambda}}{\partial \psi}(\langle \vec{B}^2\rangle  - \langle \vec{E}^2\rangle)=0,
\label{fullsystem}
\end{eqnarray}
where, as in Eq. (\ref{SR1}), the reduced Planck mass  $\overline{M}_{\mathrm{P}} = 1/\sqrt{8\pi G} = M_{\mathrm{P}}/\sqrt{8\pi}$ has been introduced. In Eq. (\ref{fullsystem}) $W(\psi)$ denotes the potential of $\psi$, $V(\varphi)$ 
(as in Eq. (\ref{SR1})) is the inflaton potential while ${\mathcal H} = a'/a$. 

The electric and magnetic  spectra  obtained in the present section assume that the system 
of governing equations are invariant under the electromagnetic duality. However, in the transition regime 
from inflation to radiation charged particles are created. Therefore the conductivity switches on 
from zero to a finite (large) value and the duality symmetry is explicitly broken.

The analysis leading to Eqs. (\ref{dual45})--(\ref{dual46})  and to Eqs. (\ref{dual47})--(\ref{dual48})
is purely kinematical.This means that a putative form of the gauge coupling has been assumed by completely 
disregarding how the dynamical scenario could be realized.  Any limit derived on the spectral 
index just highlights the necessity of a more complete description which was attempted, incidentally, in 
\cite{variation1} (see also \cite{mg5} for the analysis of electromagnetic backreaction effects 
in a different but related context).  Conversely, the (qualitative) conclusions of \cite{rep4} only demonstrate the inadequacy of an approximation for a limited range of the parameter space of the toy model under discussion.

In the case of pre-big bang models \cite{pbb} the situation is different from what has been described so far and the bounds discussed above do not arise. For instance, in the toy model used in \cite{mg5} 
the evolution of the gauge coupling and of the geometry is regular. The asymptotic evolution can be written, 
in the Einstein frame, as:
\begin{eqnarray}
&& a(\tau) \simeq a_{-} \sqrt{ -\frac{\tau}{2 \tau_0}}, ~~~~~~~~ a_{-} = e^{-\varphi_0/2} \sqrt{\frac{2(\sqrt{3} +1)}{\sqrt{3}}},
\nonumber\\
&& \varphi_{-} = \varphi_0 - \ln{2} - \sqrt{3} \ln{\biggl(\frac{\sqrt{3} +1}{\sqrt{3}}\biggr)} 
- \sqrt{3} \ln{\biggl(- \frac{\tau}{2 \tau_0}\biggr)},
\nonumber\\
&& {\cal H}_{-} = \frac{1}{2\tau}, ~~~~~~~~~~~~~~~\varphi_{-}' = - \frac{\sqrt{3}}{\tau}.
\label{solminus}
\end{eqnarray}
 for $\tau \to -\infty$. The gauge coupling will then be given, in this context, by $e^{\varphi_{-}/2}$ which implies $\nu = (1 - \sqrt{3})/2$ in the equations for $\Omega_{Y}(\tau,\nu)$. In this case the critical density bound is never violated since, to leading order in $x<1$ 
 \begin{equation}
 \Omega_{Y}(x,\nu) \simeq \frac{H^2}{M_{\mathrm{P}}^2} x^{4 - \sqrt{3}}, \qquad \frac{H^2}{M_{\mathrm{P}}} = \frac{{\mathcal H}^2
}{a^2 M_{\mathrm{P}}^2} \ll 1.  
\label{PBB}
\end{equation}
In this case the result seems to be different from the qualitative estimates of \cite{rep4} and in agreement 
with former results \cite{pbb}. Similar estimates can be carried on in the so-called string phase of the model \cite{pbb}.  Back-reaction effects 
of the gauge fields can be relevant, in bouncing models of pre-big bang type, but in the opposite regime. The amplified quantum fluctuations can re-enter the Hubble radius when the geometry is still dominated by the dilaton. In this case the energy density of the amplified fluctuations decreases more slowly than the background and the  result can be a kind of gravitational reheating which has been studied, in detail, in \cite{mg5}.

\renewcommand{\theequation}{3.\arabic{equation}}
\setcounter{equation}{0}
\section{Breaking of the duality symmetry}
\label{sec3}
The transition from inflation to radiation can be smoothly parametrized in terms of the scale factor of a conformally flat Friedmann-Robertson-Walker background 
\begin{equation}
a(\tau) = \tau + \sqrt{\tau^2 + \tau_{1}^2}.
\label{cond1}
\end{equation}
From Eq. (\ref{cond1}) it follows that 
\begin{equation}
\lim_{\tau \ll - \tau_{1}} a(\tau) \simeq 2\biggl(-\frac{\tau_{1}}{\tau}\biggr), \qquad 
\lim_{\tau \gg \tau_{1} } a(\tau) \simeq \biggl(\frac{\tau}{\tau_{1}}\biggr),
\label{cond2}
\end{equation}
implying that for $\tau\ll -\tau_{1}$ the geometry is of quasi-de Sitter type while 
for $\tau\gg \tau_{1}$ the geometry is dominated by radiation. The transition regime 
is associated with the scale $\tau_{1}$.
The smoothness of the transition is necessary if the conductivity has to be 
included consistently. 
In the presence of an Ohmic current the evolution equations discussed in  Eqs. (\ref{dual1}) 
are modified and they become \footnote{For practical reasons 
we shall use hereby Gaussian units with the $4\pi$ factor in front of the 
current.}
\begin{eqnarray}
&&\vec{\nabla} \times (\sqrt{\lambda} \vec{{B}}) = \frac{\partial}{\partial\tau} [  \sqrt{\lambda} \vec{E}] + 4\pi \vec{J},
\label{cond3}\\
&& \frac{\partial}{\partial \tau} \biggl[ \frac{ \vec{ B}}{\sqrt{\lambda}}\biggr]
+ \vec{\nabla}\times \biggl[ \frac{ \vec{ E}}{\sqrt{\lambda}}\biggr]  =0,
\label{cond4}
\end{eqnarray}
where $\vec{J}$ is the Ohmic current arising because of the interaction of the electromagnetic field with the 
ambient plasma.  As pointed out in \cite{variation1} the hypercharge 
field projects on the electromagnetic field as ${\mathcal A}_{i}^{\mathrm{em}} = \cos{\theta_{\mathrm{w}}} {\mathcal Y}_{i}$. In what 
follows, however, we will simply assume the presence of a relativistic electromagnetic plasma with positively and negatively charged species
whose masses are much smaller than the corresponding kinetic 
temperatures. 
Defining as $n_{\pm} = a^3 \tilde{n}_{\pm}$ the comoving concentrations 
of positively and negatively charged species we shall assume that the Universe 
during and after inflation is globally neutral, i.e. $n_{+} = n_{-}$ implying that 
$\vec{\nabla}\cdot\vec{E}=0$.  The Ohmic current can be written, according to the conventions adopted above,
\begin{equation}
\vec{J} = a^3 \vec{j} = a^3 \sigma_{\mathrm{c}} \vec{{\mathcal E}} = \frac{\sigma}{\sqrt{\lambda}} \vec{E}, \qquad 
\sigma = \sigma_{\mathrm{c}} a.
\label{cond4a}
\end{equation}
Note that $\sigma = a \sigma_{\mathrm{c}}$ is the rescaled conductivity which can be 
computed for a relativistic plasma with evolving gauge coupling. It is also relevant to remark that 
$\vec{J}$ scales with $1/\sqrt{\lambda}$: this occurrence can be intuitively understood by recalling that, when the gauge 
coupling is not dynamical, the current is proportional to the electric charge, i.e. $1/\sqrt{\lambda}$ in our notations. 

The conductivity is vanishingly small during inflation and it jumps to a finite value after the inflation. 
Similarly the gauge coupling will evolve dynamically during inflation and it will eventually freeze during the subsequent phases. An  interpolating expression for $\lambda(\tau)$ can be written as 
\begin{equation}
e^2 \lambda(z) = \biggl(\frac{2 \sqrt{z^2 +1}}{\sqrt{z^2 + 1} + z} \biggr)^{\frac{3\alpha}{2}},\qquad z = \frac{\tau}{\tau_{1}}.
\label{SM1a}
\end{equation}
where $\alpha = (1 - 2 \nu)/3$. From the algebraic form of Eq. (\ref{SM1a})  it follows that 
\begin{eqnarray}
 \lim_{z \gg 1} \lambda(z) = \frac{1}{e^2}\qquad \lim_{z \ll -1 } \lambda(z) = \frac{(-z)^{3\alpha}}{e^2}.
\label{SM1b}
\end{eqnarray}
In the same interval of (rescaled) conformal time the conductivity must jump from zero to a finite value
which coincides with the temperature and this behaviour can be modeled as 
 \begin{equation}
 \sigma_{\mathrm{c}}(z) = \frac{4\pi T_{\mathrm{rh}}}{e^2} \theta(z),\qquad \theta(z) = \frac{1}{2^{\beta}} 
 \biggl( 1 + \frac{z}{\sqrt{z^2 + 1}}\biggr)^{\beta}.
 \label{SM5}
 \end{equation}
 In Eq. (\ref{SM5}) $T_{\mathrm{rh}}$ is the reheating temperature.
 By increasing $\beta$ the transition becomes sharper. Indeed as  
$\beta$ increases $\sigma_{\mathrm{c}}$ vanishes more accurately 
during the inflationary stage of expansion.
   
In the presence of the conductivity the evolution equations of the mode functions are modified as follows 
\begin{eqnarray}
&& g_{k}' = - k^2 f_{k} - \frac{4\pi \sigma }{\lambda} g_{k} - \frac{(\sqrt{\lambda})'}{\sqrt{\lambda}} g_{k},
\label{MF1}\\
&& f_{k}' = \frac{(\sqrt{\lambda})'}{\sqrt{\lambda}} f_{k} + g_{k}.
\label{MF2}
\end{eqnarray}
The evolution equation of the mode functions in the conformal time coordinate $\tau$ becomes simply:
 \begin{equation}
 f_{k}'' + \frac{4\pi\sigma}{\lambda} f_{k}'+ \biggl\{k^2 - \biggl[ \frac{(\sqrt{\lambda})''}{\sqrt{\lambda}} + 
 \frac{4\pi \sigma}{\lambda} \frac{(\sqrt{\lambda})'}{\sqrt{\lambda}}\biggr]\biggr\} f_{k}=0.
 \end{equation}
 When we pass from $\tau$ to $z=\tau/\tau_{1}$ the terms containing $\sigma$, for power counting 
 reasons get multiplied by $\tau_{1} = 1/{\mathcal H}_{1} = 1/(a_1 H_1)$. The relevant 
 combination can then be written  
 \begin{equation}
 \frac{4\pi \sigma \tau_{1}}{\lambda} = \frac{4\pi \sigma_{\mathrm{c}}(z) \tau_{1} a(z)}{\lambda(z)} = \frac{16\pi^2 T_{\mathrm{rh}}}{H_{1}}\frac{a(z) \theta(z)}{e^2 \lambda(z)}.
 \end{equation}
 where we used the fact that $\sigma(z) = \sigma_{\mathrm{c}}(z) a(z)$ and that 
 \begin{equation}
 H_{1}^2 \overline{M}_{\mathrm{P}}^2 = \frac{\pi^2}{90} N_{\mathrm{eff}} T_{\mathrm{rh}}^4,\qquad T_{\mathrm{rh}} = \biggl(\frac{90}{\pi^2 N_{\mathrm{eff}}}\biggr)^{1/4} \sqrt{H_{1} \overline{M}_{\mathrm{P}}},
 \end{equation}
where $N_{\mathrm{eff}}$ is the effective number of relativistic species which will be assumed to coincide, for 
simplicity, with $106.75$ as in the standard electroweak model.

Conductivity effects might also play a role at the onset of the inflationary evolution.
The maximal number of inflationary e-folds accessible
to our observation correspond to the minimal number of e-folds 
required to relax the problems of the standard cosmological model.
If the case of a standard post-inflationary thermal history 
$N_{\mathrm{max}} = 67.951 + 0.25 
\ln{(\pi \epsilon {\mathcal A}_{{\mathcal R}})}$. For our fiducial set of parameters 
$N_{\mathrm{max}} = 62.3$. This estimate is consistent with 
the result obtained in \cite{LL} for the number of e-foldings 
before the end of inflation at which observable perturbations 
were generated. A deviation from the standard post-inflationary 
history (with the plausible addition of a stiff epoch) 
can reduce $N_{\mathrm{max}}$ even by $10$ e-folds. 
If the number of inflationary e-folds is ${\mathcal O}(63)$ the 
initial state of the electromagnetic perturbations 
is sensitive to the pre-inflationary expansion which is unknown and which could have 
contained a relativistic plasma. In the latter case the electric degrees of freedom 
are already suppressed when the initial conditions of the mode functions are set. 

\renewcommand{\theequation}{4.\arabic{equation}}
\setcounter{equation}{0}
\section{Phenomenological considerations}
\label{sec4}
In the case of conventional inflationary 
models the evolution of the geometry starts, typically, from a highly curved stage of expansion. 
 If the duration of inflation is not minimal the Hubble rate at the onset of inflation can be close to the Planck scale. Assuming $\epsilon =0.02$ 
from the observational data  and supposing, for instance, that inflation lasted a number of e-folds six times larger than  the minimal amount (i.e., overall, $N_{\mathrm{efolds}}\sim 360$ or even larger) the Hubble rate at the onset of inflation could have been  ${\mathcal O}(10^{-2.3}) M_{\mathrm{P}}$ or larger. Since the inflationary evolution 
commences in a regime of strong gravitational coupling,  it is not unreasonable that also the gauge coupling could be  strong at the onset of the dynamical evolution. 
Consider, as an example, the case $\alpha = - 4/3$ and $\nu = 5/2$. The evolution equations of the electric fields can be integrated numerically by imposing quantum mechanical initial conditions 
at the time $\tau_{\mathrm{i}}$. The choice of quantum mechanical initial conditions 
is appropriate due to the largeness of $N_{\mathrm{efolds}}$. On the contrary, 
if the pre-inflationary evolution is dominated by a relativistic plasma   
the minimal number of e-folds would not 
guarantee the plausibility of quantum mechanical initial conditions 
at $\tau_{\mathrm{i}}$.
\begin{figure}[!ht]
\centering
\includegraphics[height=6.7cm]{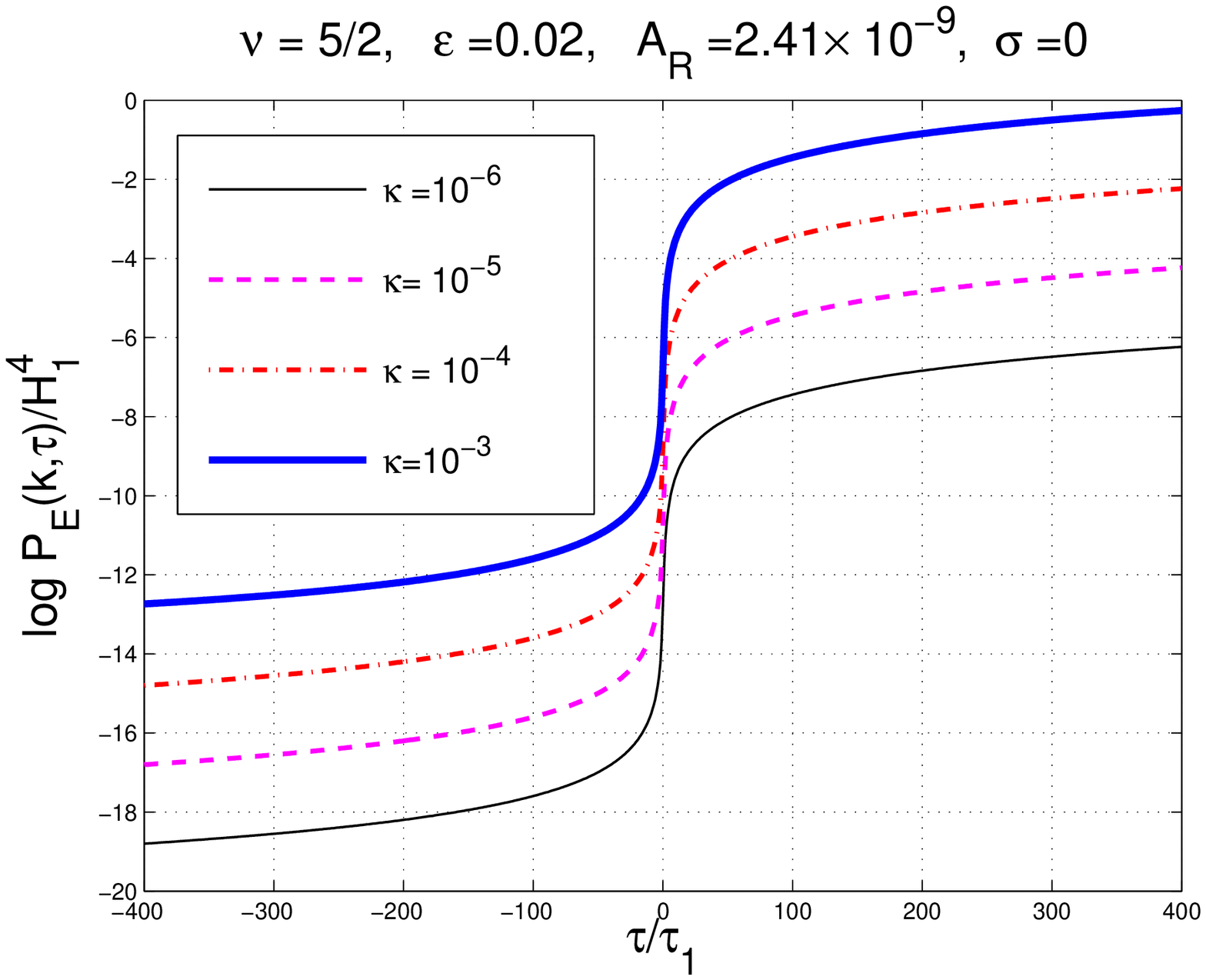}
\includegraphics[height=6.7cm]{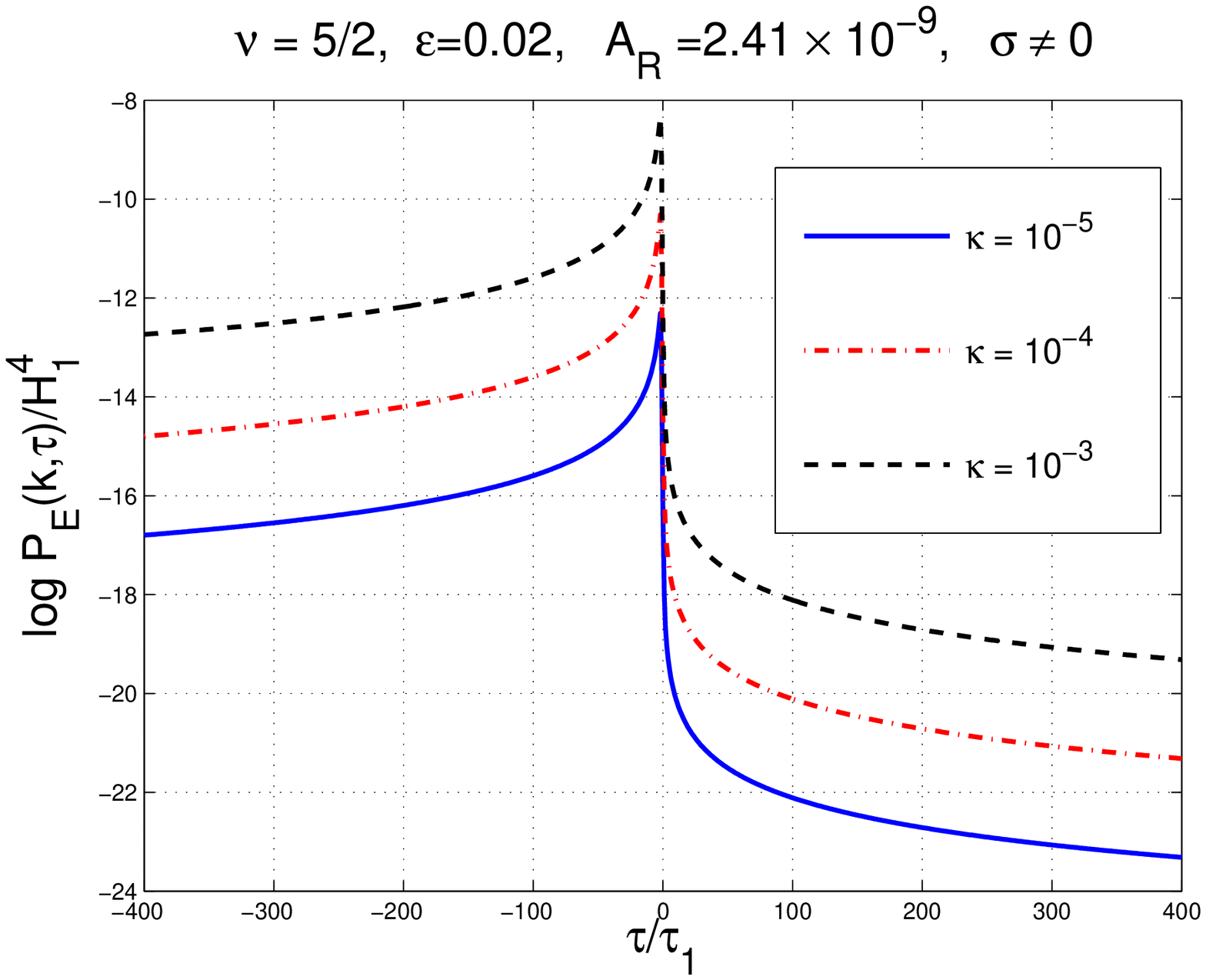}
\caption{The electric power spectrum with (i.e. $\sigma \neq 0$) and without 
(i.e. $\sigma =0$) conductivity.} 
\label{FIGURE1}
\end{figure}
It is now practical to introduce
the rescaled wavenumber $\kappa = k \tau_{1}$ where 
$\tau_{1}$ sets the scale of the transition from the inflationary phase 
to the  radiation dominated phase  (see Eq. (\ref{cond1})). At the onset of the numerical integration 
all the wavelengths are all inside the Hubble radius, i.e. $\kappa z_{\mathrm{i
}} >1 $ where $z_{\mathrm{i}} = \tau_{\mathrm{i}}/\tau_{1}$. 
In Fig. \ref{FIGURE1} the result of the numerical integration is illustrated 
for the case $\nu=5/2$. 
In the plot at the left the conductivity vanishes 
and, as expected, the spectrum of the electric field scales according to the 
analytical result, i.e. ${\mathcal P}_{\mathrm{E}} \simeq \kappa^2$.
In the plot at the right the conductivity is included and the electric 
power spectrum is exponentially suppressed. The exponential suppression 
does depend upon the evolution of the conductivity across the matter 
radiation transition. In Fig. \ref{FIGURE1} the plot at the right assumes 
the parametrization of Eq. (\ref{SM5}) with $\beta \gg 1$ (in practice $\beta =8$). 
The latter choice guarantees that during the inflationary phase $\sigma$ 
vanishes rather accurately. In the opposite case (i.e. $\beta \simeq {\mathcal O}(1)$) the presence of the conductivity can modify both the amplitude 
and the slope of the spectra. In the simplified model discussed 
here the critical requirement is therefore that $\sigma/\lambda$ 
must vanish much faster than, say, $\lambda''/\lambda$ for 
$\tau \ll -\tau_{1}$.
In Fig. \ref{FIGURE2}, Eq. (\ref{dual22}) is numerically illustrated. The result 
of Fig. \ref{FIGURE2} implies that the evolution of the canonical operators 
is such that $[\hat{y}_{i}, \hat{\pi}_{j}] \to 0$ for $\tau \geq \tau_{1}$.
Because of the presence of the conductivity the Wronskian is driven to zero. This 
means that out of the two solutions of the system only one survives, i.e. the one 
related to the magnetic part.  The vanishing of the Wronskian signals 
the transition to the classical dynamics where the magnetic field operators become 
Gaussian random fields, i.e., 
\begin{equation}
B_{i}(\vec{x},\tau) = \frac{1}{(2\pi)^{3/2}} \int d^{3} k \, B_{i}(\vec{k},\tau),
\qquad \langle B_{i}(\vec{k},\tau) B_{j}(\vec{p},\tau) \rangle = \frac{2\pi^2}{k^3} {\mathcal P}_{\mathrm{B}}(k) \,P_{ij}(k)\,\delta^{(3)}(\vec{k} + \vec{p}),
\label{GRF}
\end{equation}
\begin{figure}[!ht]
\centering
\includegraphics[height=6.7cm]{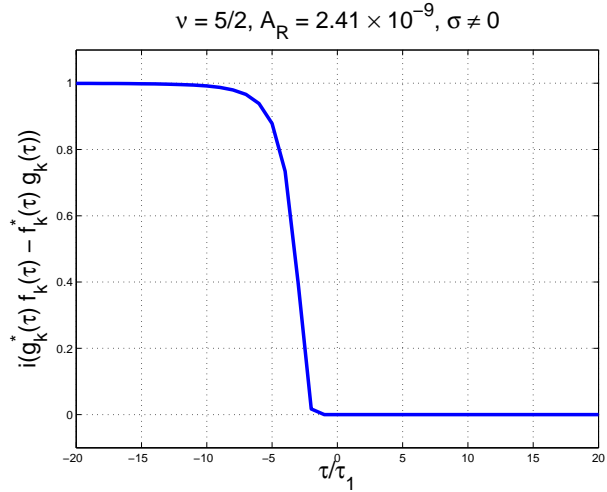}
\caption[a]{The vanishing of the Wronskian is illustrated.}
\label{FIGURE2}
\end{figure}
In Fig. \ref{FIGURE3} the spectra of the magnetic fields are illustrated 
always in the case of decreasing gauge coupling. It is practical
to assign the magnetic power spectra with the same conventions 
employed in the case of   the power spectrum of curvature perturbations 
(see Eq. (\ref{scten})):
\begin{equation}
{\mathcal P}_{\mathrm{B}}(k) = {\mathcal A}_{\mathrm{B}} \biggl(\frac{k}{k_{\mathrm{L}}}\biggr)^{n_{\mathrm{B}} -1}, 
\label{PB1}
\end{equation}
where $k_{\mathrm{L}} = 1\, \mathrm{Mpc}^{-1}$ is the magnetic pivot 
scale and ${\mathcal A}_{\mathrm{B}}$ is the spectral amplitude at $k_{\mathrm{L}}$. In Fig. \ref{FIGURE3} the cases $n_{\mathrm{B}} =1$ 
and $n_{\mathrm{B}} = 1.5$ are explicitly illustrated. When the 
transition to radiation is delayed (in the case of a small $\beta$) 
the magnetic fields are more suppressed than in the case 
of $\beta \gg 1$ (which is the one illustrated both in Figs. \ref{FIGURE1} and \ref{FIGURE3})). The results are most easily normalized 
in terms of $H_{1}^4$ where $H_{1} \equiv {\mathcal H}(0)$
since ${\mathcal H}(\tau) = 1/\sqrt{\tau^2 + \tau_{1}^2}$. 
In the case of flat spectrum the amplitude ${\mathcal A}_{\mathrm{B}}$ is 
solely controlled by the amplitude of the curvature perturbations and it can be written as 
\begin{equation}
\frac{{\mathcal A}_{\mathrm{B}}}{\rho_{\gamma}(\tau_{0})} = 18.49 
\,  \epsilon\, {\mathcal A}_{\mathcal R} \biggl(\frac{T_{\gamma\,0}}{2.725\, \mathrm{K}}\biggr)^{-4} ,\qquad \sqrt{\rho_{\gamma}(\tau_{0}
)} = 2.29 \, \mu \mathrm{G}
\label{PB2}
\end{equation}
where $\rho_{\gamma}(\tau_{0}) = \pi^2 T_{\gamma 0}^4/15$ is the energy 
density of the CMB photons. Equation (\ref{PB2}) implies that
$\sqrt{A_{\mathrm{B}}} = 6.840\times 10^{-11} \, G = 0.068 \, \mathrm{nG}
$ which agrees with former estimates \cite{variation1}. To convert 
the various results in physical units it is useful to recall, as mentioned
in Eq. (\ref{PB2}),  that the energy density of the CMB photons does 
correspond to a putative magnetic field of about $3\, \mu \mathrm{G}$ 
($1\,\mu G= 10^{-6}\, \mathrm{G}$) which is, in turn, comparable 
with the energy density of the cosmic rays as already argued many years ago\cite{fermi}. 
\begin{figure}[!ht]
\centering
\includegraphics[height=6.7cm]{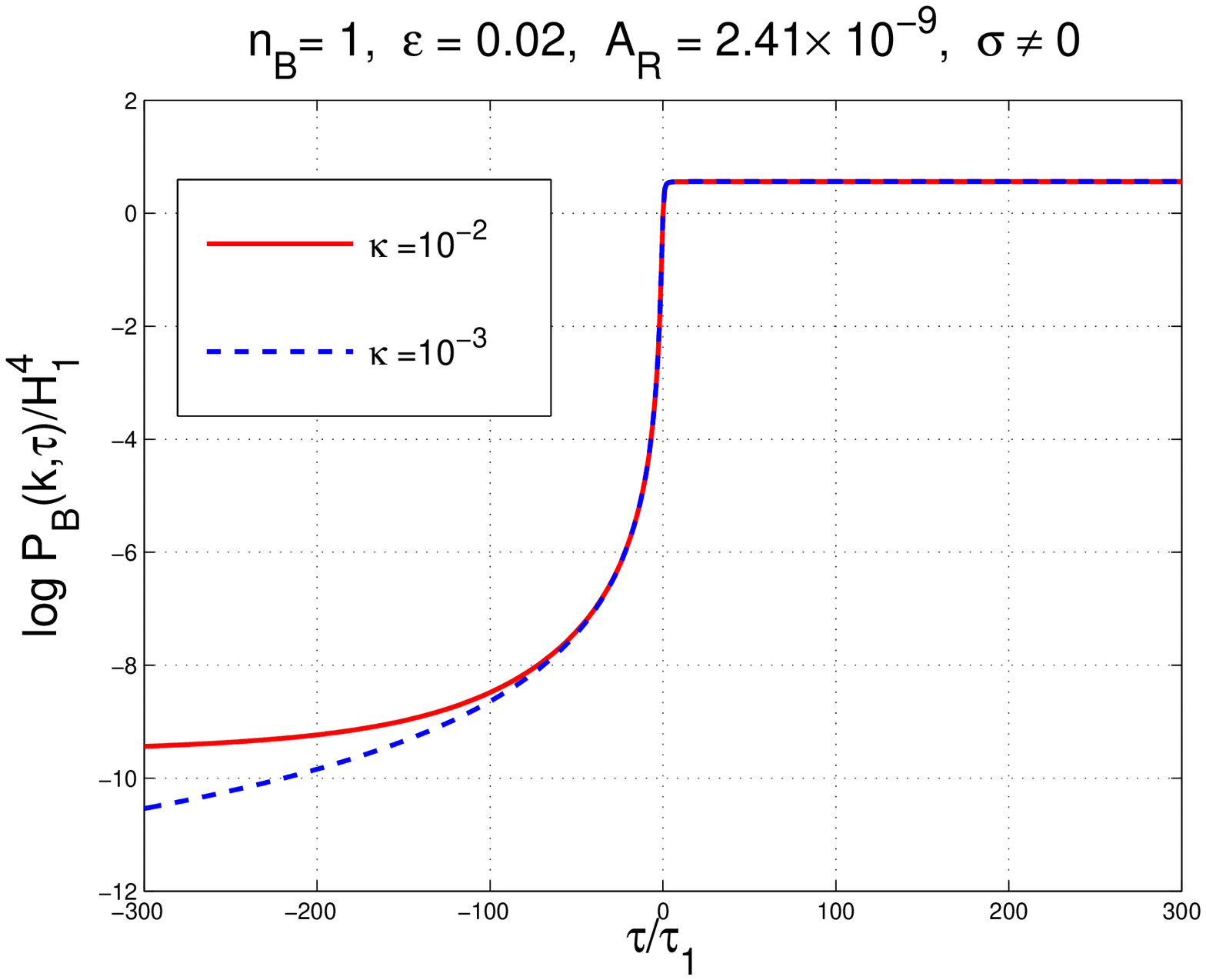}
\includegraphics[height=6.7cm]{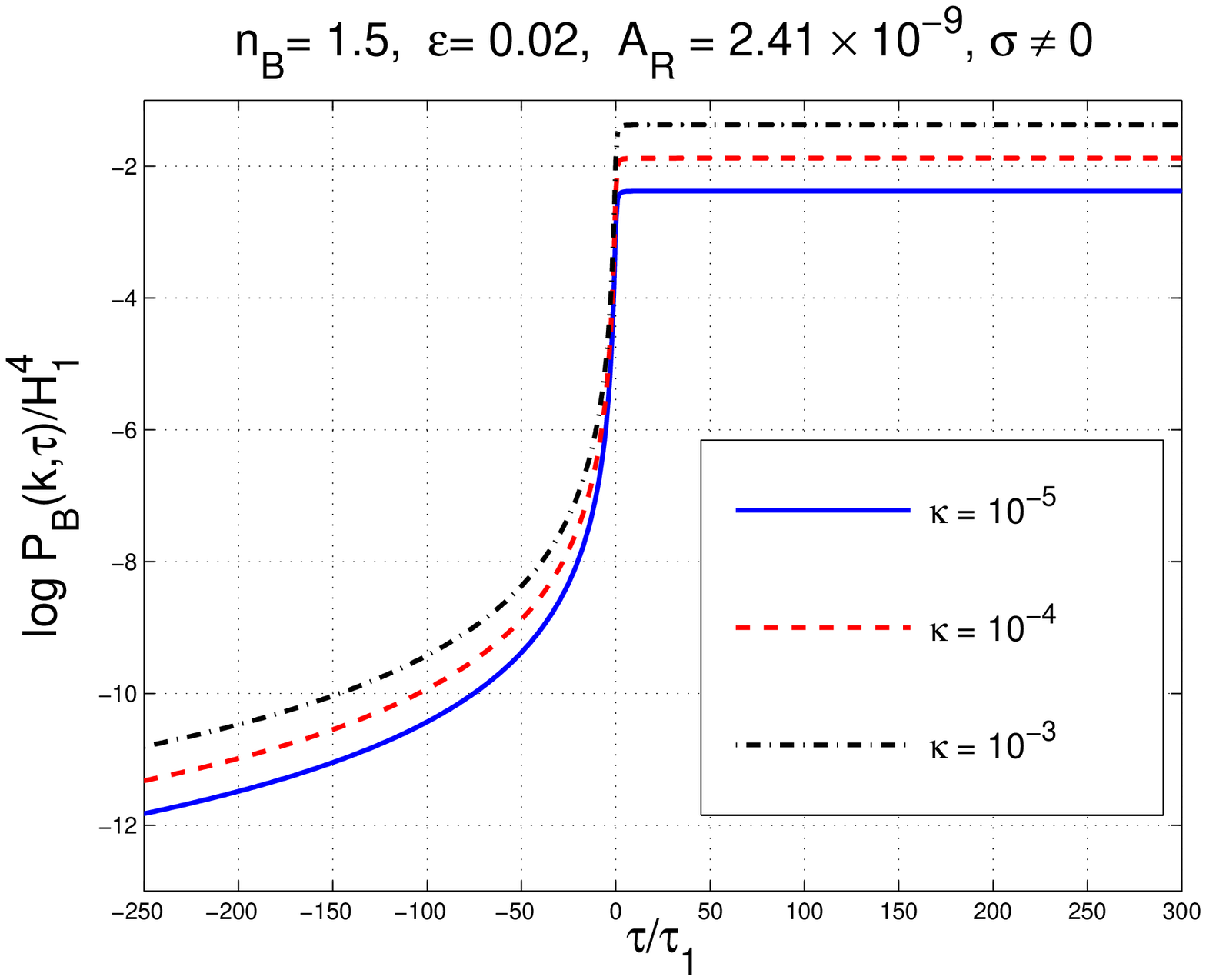}
\caption[a]{The magnetic field spectra for two different values of the spectral index 
(i.e. $n_{\mathrm{B}} =1$ and $n_{\mathrm{B}} = 1.5$) and for 
different wavenumbers.}
\label{FIGURE3}
\end{figure}
In the case $n_{\mathrm{B}} > 1$ the analog of Eq. (\ref{PB2}) 
can be written as 
\begin{equation}
\frac{{\mathcal A}_{\mathrm{B}}}{\rho_{\gamma}(\tau_{0})} = 
10^{F(n_{\mathrm{B}},\epsilon)} \biggl({\mathcal A}_{{\mathcal R}} \epsilon\biggr)^{(5 - n_{\mathrm{B}})/4}\,\,
\biggl(\frac{H_{\mathrm{r}}}{H_{\mathrm{i}}}\biggr)^{(1 - 2\mu)(n_{\mathrm{B}} - 5)/2},
\label{PB3}
\end{equation}
where $H_{\mathrm{i}}$ denotes the Hubble rate at the end of 
inflation, $H_{\mathrm{r}}$ is the Hubble rate at the onset 
of the radiation-dominated phase and $\mu$ parametrizes 
the (average) exponent of the scale factor between $H_{\mathrm{i}}$ 
and $H_{\mathrm{r}}$. The function $F(n_{\mathrm{B}},\epsilon)$ is given by:
\begin{equation}
27.362 - n_{\mathrm{B}}\,
\left[ 26.307 + 0.434\,\ln (1 - \epsilon) \right]  + 2.171\,\ln{(1 - \epsilon)} + 
  0.868\,\ln{[\Gamma( 3 - n_{\mathrm{B}}/2)]}.
\label{PB4}
\end{equation}
It can be readily checked that in the case $H_{\mathrm{i}}= H_{\mathrm{r}}$ 
Eqs. (\ref{PB3}) and (\ref{PB4}) reproduce the result of Eq. (\ref{PB2}) 
when $n_{\mathrm{B}} =1$ and for the typical values of ${\mathcal A}_{{\mathcal R}}$ and $\epsilon$. Equations (\ref{PB3}) and (\ref{PB4})
hold, both in the case of decreasing and in the case of increasing
gauge coupling. However the relation of $\nu$ to $n_{\mathrm{B}}$ changes 
slightly in the different cases. If $\nu > 1/2$ we shall have 
$\nu = (3 - n_{\mathrm{B}}/2)$ with $n_{\mathrm{B}} < 5$. If 
$\nu < 1/2$ we have to distinguish two cases. If $0 < \nu < 1/2$ 
we shall have that $\nu = (3 - n_{\mathrm{B}}/2)$ (with $5 < n_{\mathrm{B}} <6$); if $\nu < 0$ then $\nu = (n_{\mathrm{B}} -6)/2$.

If the duration of inflation is close to minimal the initial conditions of the electromagnetic fluctuations depend upon the pre-inflationary evolution much more than the initial conditions of, for instance, the scalar and the tensor modes of the geometry. Suppose, for instance, that the pre-inflationary expansion rate was dominated by a relativistic plasma. Suppose, furthermore, that inflation 
lasts around $63$ e-folds implying that, at the onset of inflation, the Hubble rate was indeed 
${\mathcal O}(10^{-5})$. By assuming a sudden transition from the pre-inflationary 
regime to the inflationary phase, we can estimate the upper limit on the temperature 
which is given by $T^4 \sim 90 H^2 \overline{M}_{\mathrm{P}}^2 /(\pi^2 N_{\mathrm{eff}})$ where 
$N_{\mathrm{eff}}$ is the effective number of relativistic species. Since the gauge coupling is small, no 
matter how small the temperature is, the conductivity will be very large since, approximately, as we saw 
$\sigma_{\mathrm{c}}\simeq T/\alpha_{Y}$ where $\alpha_{Y} = g_{Y}^2/(4\pi)$. In this situation the electric fields will be suppressed in comparison with the magnetic fields and the initial conditions for the mode functions will have to reflect this occurrence. 

\begin{figure}[!ht]
\centering
\includegraphics[height=6.7cm]{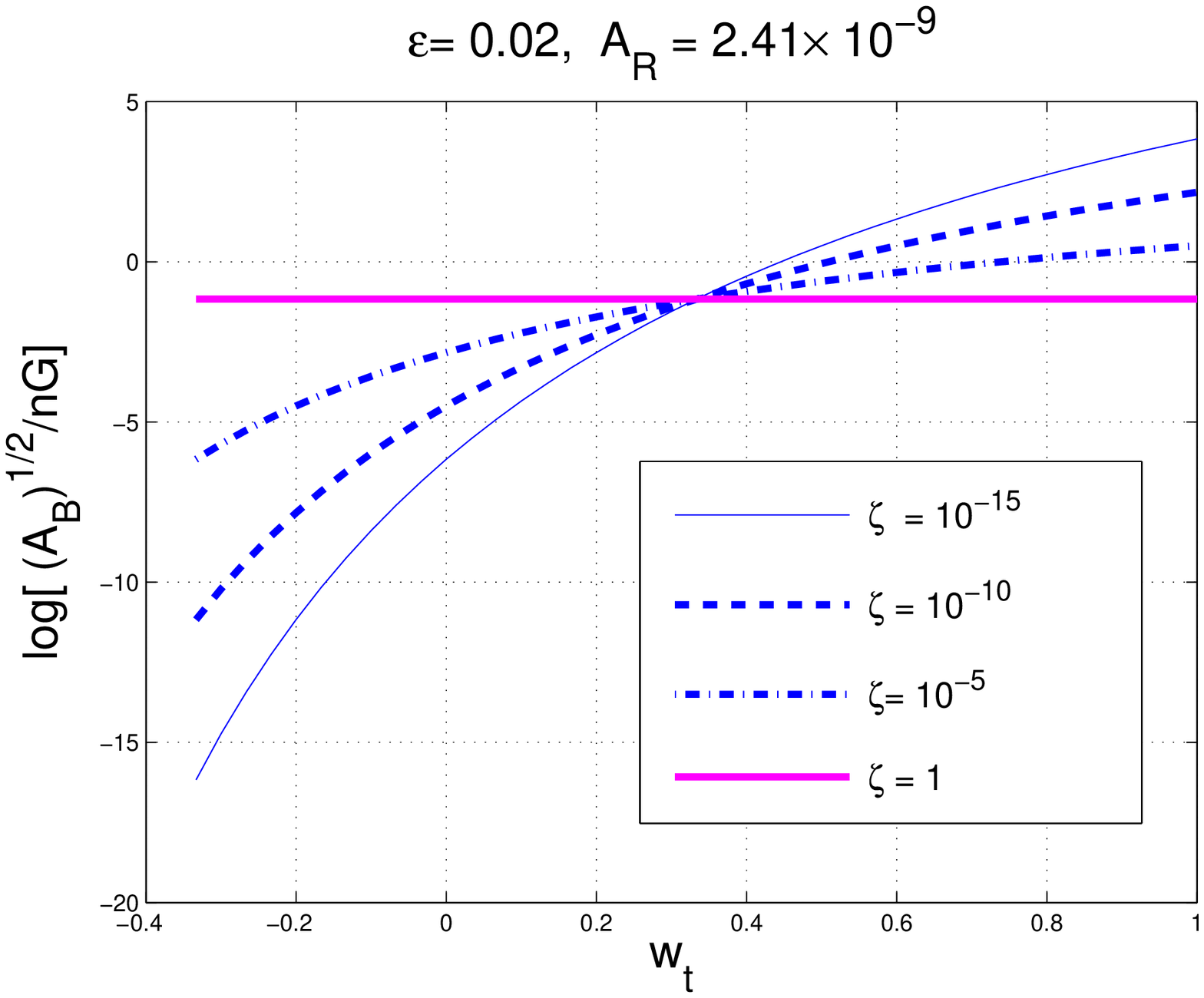}
\includegraphics[height=6.7cm]{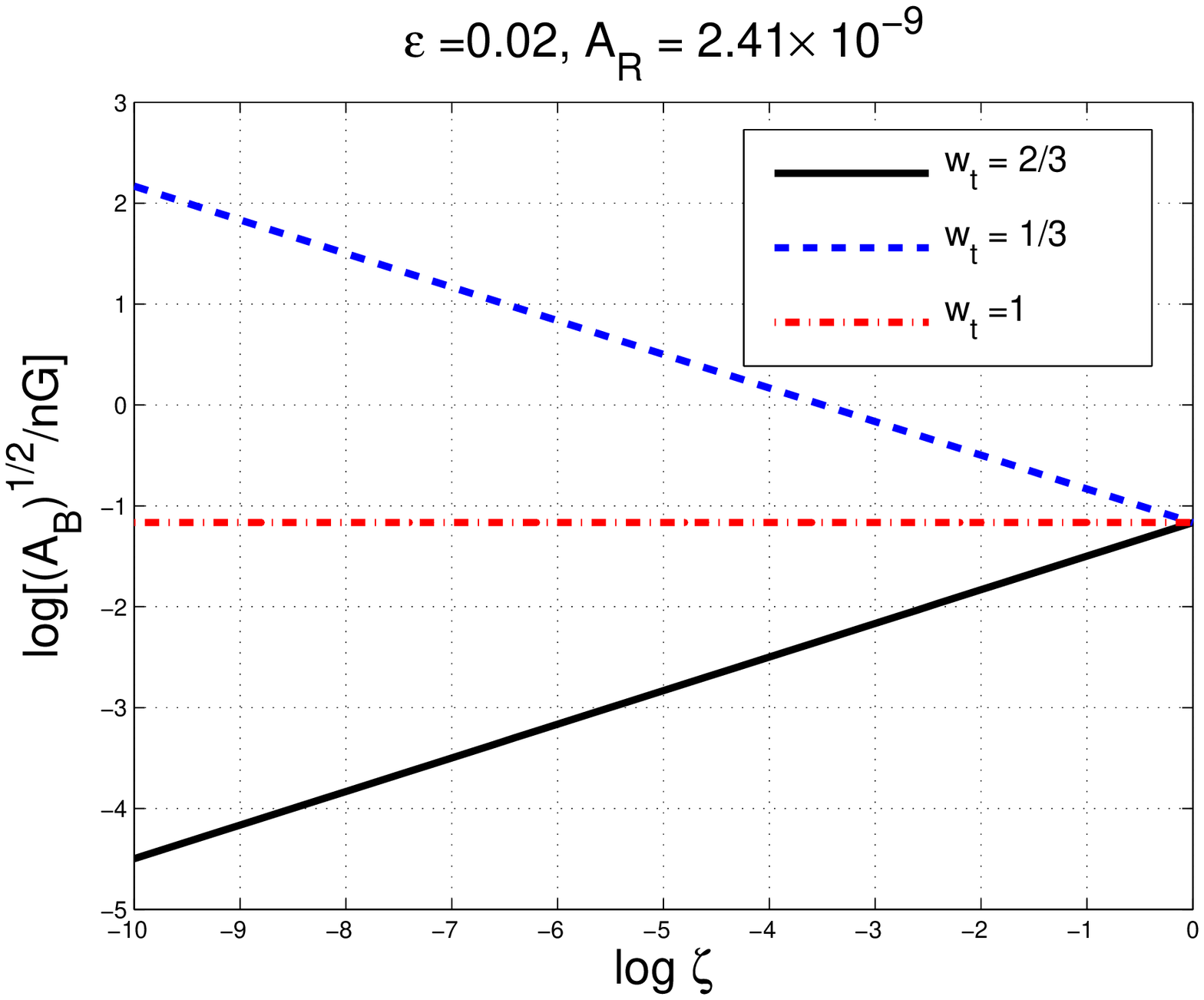}
\caption[a]{The magnitude of the scale-invariant amplitude 
is illustrated as a function of $w_{\mathrm{t}}$ and as a 
function of $\zeta$.}
\label{FIGURE4}
\end{figure}

It is now the moment of assessing the orders of magnitude 
of the magnetic fields obtained within this class of mechanisms. 
Let us first consider the case of an exactly scale-invariant spectrum (i.e. 
$n_{\mathrm{B}} = 1$) and let us measure 
$\sqrt{{\mathcal A}_{\mathrm{B}}}$ in units of nG. The amplitude of the 
scale-invariant spectrum does depend, primarily, upon $\epsilon$ and upon 
${\mathcal A}_{{\mathcal R}}$. The two horizontal lines both in the 
left and in the right plot of  Fig. \ref{FIGURE4} illustrate the minimal case 
where $\zeta = H_{\mathrm{r}}/H_{\mathrm{i}}=1$. In this situation 
the Hubble rate at the onset of the radiation epoch coincides 
with the Hubble rate at the end of inflation. It can happen 
that the onset of the radiation-dominated epoch is delayed either because 
the reheating process is prolonged or because of the presence 
of an intermediate phase. This situation is parametrized, in Fig. \ref{FIGURE4}
by the value of $\zeta < 1$ and by the barotropic index during the intermediate phase. 
If $w_{\mathrm{t}} > 1/3$ the scale-invariant amplitude increases and this occurrence 
can be appreciated both from the left and from the right plots of Fig. \ref{FIGURE4}.
In the plot at the left of Fig. \ref{FIGURE4} the amplitude is plotted as a function 
of the barotropic index of the intermediate phase. The lower limit for $w_{\mathrm{t}}$
coincides with $-1/3$ and it corresponds to a phase when the deceleration parameter 
vanishes.  The upper limit on $w_{\mathrm{t}}$ is $1$. In the case of constant barotropic index $c_{\mathrm{st}}^2 = w_{\mathrm{t}}$.
In the limiting case  $w_{\mathrm{t}} = 1 = c_{\mathrm{st}}^2$ 
and the speed of sound coincides with the speed of light. As argued 
in \cite{SpS}, barotropic indices $w_{\mathrm{t}} >1$ would not be 
compatible with causality.  Post-inflationary phases stiffer 
than radiation (i.e. $1/3< w_{\mathrm{t}} \leq 1$)  can arise in different contexts 
(see, for instance, \cite{stiff}) where $\zeta$ can be as small as $10^{-15}$.  If $w_{\mathrm{t}} < 1/3$ the conclusions are opposite: the scale invariant amplitude is always smaller than in the case when $\zeta =1$ when the transition to radiation is sudden.The results of Fig. \ref{FIGURE4} show that scale-invariant amplitudes 
ranging from $0.01$ nG to few nG cannot be theoretically excluded 
in the present framework. 

\begin{figure}[!ht]
\centering
\includegraphics[height=6.7cm]{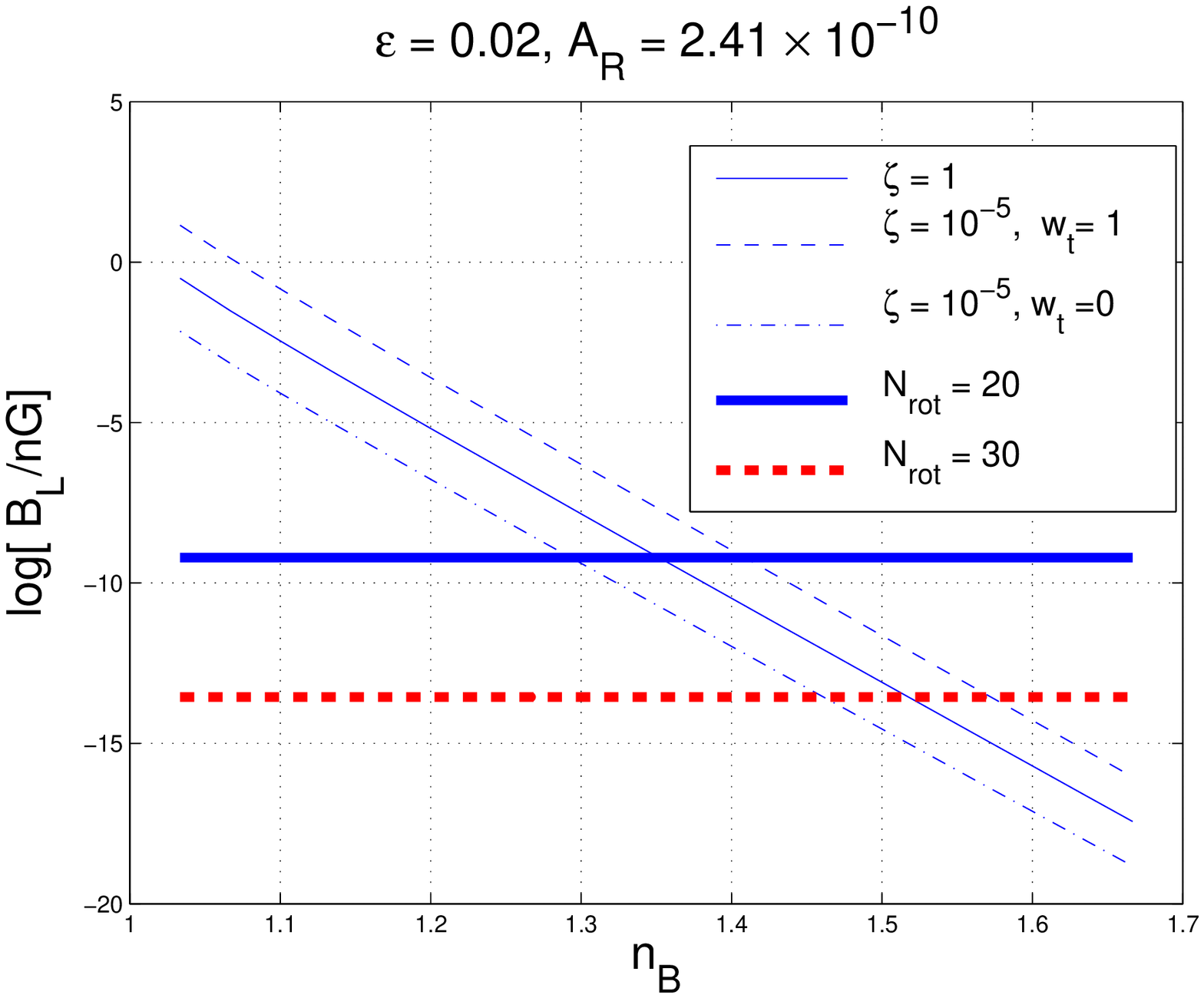}
\includegraphics[height=6.7cm]{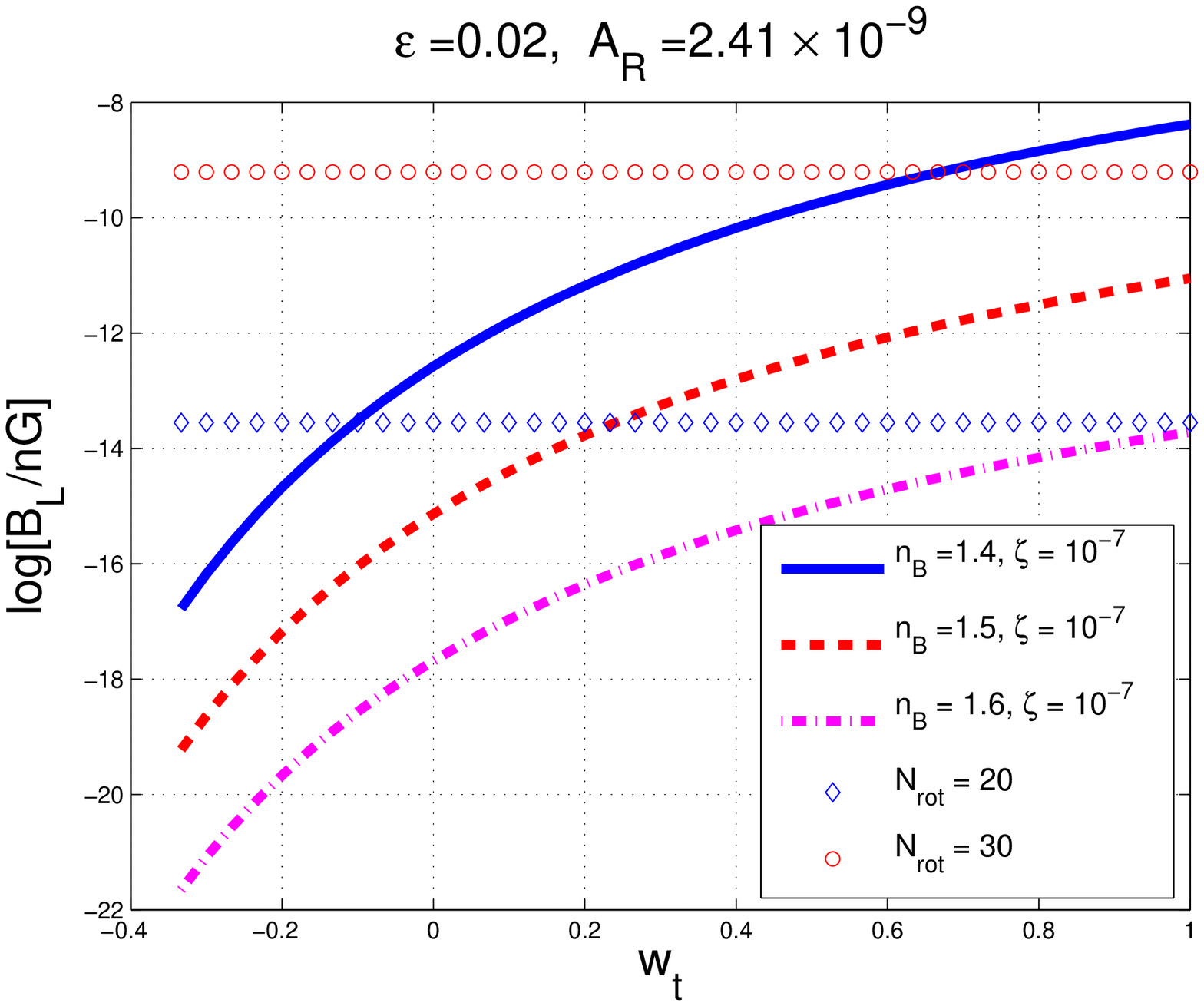}
\caption[a]{The regularized amplitude $B_{\mathrm{L}}$ is 
illustrated in the case of blue spectral indices.}
\label{FIGURE5}
\end{figure}

Let us now move to the case of blue spectral indices which are also 
the ones phenomenologically preferred at least heeding recent 
attempts aimed at constraining possible distortions of CMB anisotropies
induced by large-scale magnetic fields. 
In this case the magnetic field intensity must be regularized and 
the commonly employed strategy is to use a Gaussian window 
function so that 
\begin{equation}
B_{\mathrm{L}}^2 = \langle B_{i}(\vec{x}, \tau_{0}) B_{i}(\vec{x},\tau_{0}) \rangle 
\equiv \int \frac{d^3 k}{(2\pi)^3} \langle B_{i}(\vec{k},\tau_{0}) B_{i}(\vec{k},\tau_{0}) 
\rangle e^{- k^2 L^2},
\label{reg1}
\end{equation}
where $L= 2\pi/k_{\mathrm{L}}$. Equation (\ref{reg1}) implies 
that  
$B_{\mathrm{L}}^2 = {\mathcal A}_{\mathrm{B}} 
(2\pi)^{1 - n_{\mathrm{B}}}\, \Gamma((n_{\mathrm{B}}-1)/2)$.
 In Fig. \ref{FIGURE5} the regularized amplitude is illustrated as a function 
 of the spectral index and as a function of the barotropic index. 
 The horizontal lines in both plots illustrate the requirements 
 stemming from magnetogenesis demanding that the regularized 
 amplitude must be, at least, larger than 
 \begin{equation}
 \overline{B}_{\mathrm{L}} = 3 \,10^{3}\, e^{- N_{\mathrm{rot}}} \biggl(\frac{\rho_{\mathrm{a}}}{\rho_{\mathrm{b}}}\biggr)^{2/3} \,\, \mathrm{nG},
 \label{reg2}
 \end{equation}
 where $N_{\mathrm{rot}}$  is the number of (effective) rotations performed 
 by the galaxy since gravitational collapse; $\rho_{\mathrm{a}}$ and $\rho_{\mathrm{b}}$ 
 denote, respectively, the matter density after and before gravitational collapse. 
 The typical rotation period of a spiral galaxy is of the order of $3\times10^{8}$ yrs which should be compared with $10^{10}$ yrs, i.e. the approximate age of the galaxy.  The maximal number of rotations 
 performed by the galaxy since its origin is of the order of $N_{\mathrm{rot}}\sim 30$ (i.e. 
the ratio of the two figures mentioned in the preceding sentence). 
 This generous estimate refers to the situation where the kinetic energy of the plasma 
 is transferred to the magnetic energy by means of an extremely efficient 
 dynamo which amplifies the magnetic field by one efold for each rotation. The effective 
 number of efolds is always smaller than $30$ for various reasons. Typically it can happen that the dynamo quenches prematurely because some the higher wavenumbers  
 of the magnetic field become critical (i.e. comparable with the kinetic energy of the plasma) before the smaller ones. Other sources of quenching have been recently discussed in the literature (see, for an introduction to this topic, section 4.2 of \cite{dyn} and references therein). 
 
 There is also another source of amplification of the primordial magnetic field and it has to do with compressional 
amplification. At the time of the gravitational collapse of the 
protogalaxy the conductivity of the plasma was sufficiently high 
to justify the neglect of nonlinear corrections in the equations 
expressing the conservation of the magnetic flux and of  the 
magnetic helicity. The conservation of the magnetic flux
implies that, during the gravitational collapse, the magnetic field 
should undergo compressional amplification, i.e. the same 
kind of mechanism which is believed to be the source of the 
large magnetic fields of the pulsars.  
Right before the gravitational collapse of the protogalaxy 
the mean matter density can be taken to be of the 
order of the critical density, i.e. $\rho_{\mathrm{b}} \simeq
 \rho_{\mathrm{crit}}$. After collapse the matter density $\rho_{\mathrm{a}}$ of the galaxy is larger by, roughly, $6$ or $7$ orders of magnitude.
 Thus, after gravitational collapse, the magnetic field gets roughly  amplified by a factor $(\rho_{\mathrm{b}}/\rho_{\mathrm{a}})^{2/3} \simeq 10^{4}$ in spite of the possible occurrence of a strong or moderate dynamo action.  It is worth mentioning that the 
 scale-invariant amplitudes illustrated in Fig. \ref{FIGURE4}
 satisfy the magnetogenesis requirements even in the absence 
 of  a strong dynamo action. 

In Fig. \ref{FIGURE5} the two horizontal lines in both plots 
illustrate the requirement expressed by Eq. (\ref{reg2}) 
in the case of two different numbers of effective rotations.
In both plots of Fig. \ref{FIGURE5} the regions which are above 
the horizontal lines allow for a viable class of magnetogenesis 
scenarios.  
In the sudden reheating approximation (i.e. $\zeta = 1$) 
the left plot of Fig. \ref{FIGURE5} shows that, for instance, 
$n_{\mathrm{B}} < 1.35$ to be compatible with the 
magnetogenesis requirement when $N_{\mathrm{rot}} =20$ (see 
thin oblique line in left plot of Fig. \ref{FIGURE5}). 
A relatively short delay in the beginning of the radiation epoch 
(i.e. $\zeta \simeq 10^{-5}$) allows for $n_{\mathrm{B}} <1.4$ 
provided $w_{\mathrm{t}} > 1/3$, i.e. provided 
the sound speed during the intermediate phase is larger than $1/\sqrt{3}$. If, on the contrary, $w_{\mathrm{t}} <1/3$ the delay of the radiation results in a decrease of the magnetic field intensity and, ultimately, in a stronger bound on $n_{\mathrm{B}}$ (i.e. $
n_{\mathrm{B}} <1.3$); this features are illustrated in the left plot of Fig. \ref{FIGURE5} with the dashed and with the dot-dashed 
oblique lines. In the right plot of Fig. \ref{FIGURE5} the 
magnetogensis bounds are shown with the two horizontal 
lines (with naughts and diamonds). As written in the legend 
the delay of the radiation phase is assumed to be $\zeta \simeq 10^{-7}$ and the magnetic field intensity is plotted as a 
function of $w_{\mathrm{t}}$. 

Large-scale magnetic fields have been also observed in regular  (or Abell) clusters and statistical determinations of intra-cluster magnetic fields are available since almost ten years \cite{cl,cl2,clrev} (see also \cite{mg1} for a general perspective  on cluster magnetism).  
These statistical determinations of cluster magnetic fields 
have been made possible thanks to the combination of Faraday 
rotation measurements with the determinations of the surface
brightness of the clusters in the x-rays. The latter (satellite) determination allowed to infer the electron density and, ultimately, 
the strength of the magnetic field which turns out to be in 
the range of $500$ nG. 

The typical scale of the gravitational collapse of a cluster 
is larger (roughly by one order of magnitude) than the scale of gravitational collapse of the protogalaxy. The mean mass density 
within the Abell radius ( $\simeq 1.5 h_{0}^{-1} $ Mpc) is roughly 
$10^{3}$ larger than the critical density since  clusters are 
formed from peaks in the density field. Moreover, clusters 
rotate much less than galaxies even if it is somehow 
hard to disentangle, observationally, the global (coherent) 
rotation of the cluster from the rotation curves of the 
constituent galaxies. By assuming, for instance, $N_{\mathrm{rot}}=5$, a density gradient of $10^{3}$ and $500$ nG as final field, 
Eq. (\ref{reg2}) demands and initial seed of the order $0.15$ nG 
over a typical pivot scale $\tilde{k}_{\mathrm{L}} = k_{\mathrm{L}}/10$. This requirement, in the present context, 
holds when the spectral amplitude is quasi-flat.

In spite of the fact that it is always good to attempt accurate estimates, we must also 
admit that to have firmer predictions on the pre-decoupling magnetism it is 
necessary to model more carefully the evolutionary history of the conductivity 
between the end of inflation and the onset of the radiation-dominated epoch. This 
aspect is essential when assessing the viability of this type of scenario.  

\newpage
\renewcommand{\theequation}{5.\arabic{equation}}
\setcounter{equation}{0}
\section{Concluding remarks}
\label{sec5}
In the last decade ground based as well as satellite 
observations probed with greater accuracy the nature 
of the magnetic field of the sun. The SOHO 
\cite{soho}, TRACE \cite{trace}, GONG \cite{gong}
experiments \footnote{ SOHO stands for 
Solar and Heliospheric Observatory; 
TRACE stands for  TRansition Region and Coronal Explorer;
GONG stands for Global Oscillation Network Group.}
helped in deepening solar observations. In spite
of these direct probes, it is fair to say that 
 the 22-year magnetic cycle of the Sun cannot  be claimed 
 to be fully understood in terms of the so-called 
 $\alpha\omega$ mechanism. The GONG observations 
 give, for instance, the profile of the angular velocity of the sun 
 not only on the surface but also in the interior of the sun. In spite 
 of this valuable determination other effects may play a role so that it is fair to say 
 that the situation is not clear \cite{sun}.
 
The sun is the closest and better observed astrophysical 
object and still the features of dynamo amplification 
are under debate. In the case of large-scale magnetism 
we face a similar situation.  It seems that we will probably never be able to 
know the initial conditions of the galactic dynamo as accurately as we pretend to understand the initial 
conditions of  solar dynamos where, still, crucial puzzles remain. 

The hopes of clarifying this problem in the near future might not be so forlorn.  Indeed,
as repeatedly argued,  CMB anisotropies and polarization \cite{mga,dyn} are a powerful window 
on pre-decoupling physics and, consequently, on pre-decoupling magnetic fields. 
Waiting for more direct observational evidences it is plausible to speculate that a moderate dynamo action combined 
with compressional amplification could indeed bridge the regime of the 
initial conditions with the observed large-scale magnetic fields.  Even in this simplified 
framework important theoretical problems persist and they have to do with the way 
the Universe becomes a good conductor at the end of inflation.  The early variation 
of the gauge couplings is a potential candidate for producing large-scale 
magnetic seeds for galaxies and clusters.  However, more effort is certainly required especially in 
modeling the finite conductivity effects and the self-consistent evolution of the whole system of equations.


\newpage
\begin{thebibliography}{99}

\bibitem{detect} J. S. Hall,   Science {\bf 109}, 166 (1949); W. A. Hiltner,  Science {\bf 109}, 165 (1949); 
 L. J. Davis  and J. L. Greenstein, Astrophys. J. {\bf 114}, 206 (1951).


\bibitem{fermi} E. Fermi,  Phys. Rev. {\bf 75}, 1169 (1949); 
E. Fermi and S. Chandrasekar, Astrophys. J.  {\bf 118}, 113 (1953). 

\bibitem{alv} H. Alfv\'en, Arkiv. Mat. F. Astr., o. Fys. \textbf{29 B}, 2 (1943); 
H. Alfv\'en, Phys. Rev. {\bf 75} 1732 (1949); R. D. Richtmyer and E. Teller,   Phys. Rev. {\bf 75}, 1729 (1949).

\bibitem{mg1}  M.~Giovannini, Int.\ J.\ Mod.\ Phys.\  D {\bf 13}, 391 (2004).

\bibitem{SKA} See, for instance,  http://www.skatelescope.org.

\bibitem{mga}  M. Giovannini, Phys. Rev. D {\bf 79}, 121302 (2009);  Phys.\ Rev.\  D {\bf 79}, 103007 (2009);
  M.~Giovannini and K.~E.~Kunze, Phys.\ Rev.\  D {\bf 77}, 063003 (2008); 
   M. Giovannini, PMC Phys.\  A {\bf 1}, 5 (2007);

\bibitem{auger}  J. Abraham {\it et al.} [Pierre Auger Collaboration],
Science {\bf 318}, 938 (2007); M.~Aglietta {\it et al.}  [Pierre Auger Collaboration],
Astropart.\ Phys.\  {\bf 27}, 244 (2007); E.~Roulet  [Pierre Auger Collaboration],
  Nucl.\ Phys.\ Proc.\ Suppl.\  {\bf 190}, 169 (2009).

\bibitem{RB}  R.~Beck, Space Sci.\ Rev.\  {\bf 99}, 243 (2001);  R.~Beck and M.~Krause,
 Astron.\ Nachr.\  {\bf 326}, 414 (2005);
  T.~G.~Arshakian, R.~Beck, M.~Krause, D.~Sokolff and R.~Stepanov,  arXiv:0909.3001 [astro-ph.CO].

\bibitem{variation1} M.~Giovannini,  Phys.\ Rev.\  D {\bf 64}, 061301 (2001);   Phys.\ Lett.\  B {\bf 659}, 661 (2008).

\bibitem{variation2} R.~Emami, H.~Firouzjahi and M.~S.~Movahed,  arXiv:0908.4161 [hep-th]; 
 K.~Bamba, C.~Q.~Geng and S.~H.~Ho,
 JCAP {\bf 0811} 013 (2008); K.~Bamba and M.~Sasaki,  JCAP {\bf 0702}, 030 (2007).
 
\bibitem{ratra}  B. Ratra,  Astrophys.\, J.\, Lett.  {\bf 391}, L1 (1992).

\bibitem{pbb}  M. Gasperini, M. Giovannini, and G. Veneziano, {\it Phys. Rev. Lett.} {\bf 75}, 3796 (1995). 

\bibitem{DT}  S.~Deser and C.~Teitelboim,  Phys.\ Rev.\  D {\bf 13}, 1592 (1976).

\bibitem{LF} L.~H.~Ford,  Phys.\ Rev.\  D {\bf 31}, 704 (1985).

\bibitem{mg3} M.~Giovannini,  Class.\ Quant.\ Grav.\  {\bf 20}, 5455 (2003).

\bibitem{mg4} V.~Bozza, M.~Giovannini and G.~Veneziano,  JCAP {\bf 0305}, 001 (2003).

\bibitem{mg5} M.~Giovannini,  Class.\ Quant.\ Grav.\  {\bf 21}, 4209 (2004).

\bibitem{WMAP51} G.~Hinshaw {\it et al.}  [WMAP Collaboration],  arXiv:0803.0732 [astro-ph]; 
J.~Dunkley {\it et al.}  [WMAP Collaboration], arXiv:0803.0586 [astro-ph]; 
E.~Komatsu {\it et al.}  [WMAP Collaboration],  arXiv:0803.0547 [astro-ph].

\bibitem{LL} A.~R.~Liddle and S.~M.~Leach,  Phys.\ Rev.\  D {\bf 68}, 103503 (2003)

\bibitem{abr1}  M. Abramowitz and I. A. Stegun, {\it Handbook of Mathematical Functions} (Dover, New York, 1972).

\bibitem{abr2}  A. Erdelyi, W. Magnus, F. Obehettinger, and F. Tricomi,  {\it Higher Trascendental Functions} (Mc Graw-Hill, New York, 1953).  

\bibitem{rep4} V.~Demozzi, V.~Mukhanov and H.~Rubinstein,  JCAP {\bf 0908}, 025 (2009).  

\bibitem{SpS}   G.~Ellis, R.~Maartens and M.~A.~H.~MacCallum, Gen.\ Rel.\ Grav.\  {\bf 39}, 1651 (2007).

\bibitem{stiff} B. Spokoiny, Phys. Lett. B {\bf 315}, 40 (1993); L. H. Ford, Phys. Rev. D {\bf 35}, 2955 (1987);
 M.~Giovannini,  Phys.\ Rev.\  D {\bf 58}, 083504 (1998);  Phys.\ Rev.\  D {\bf 60}, 123511 (1999).

\bibitem{dyn}  M.~Giovannini,  Class.\ Quant.\ Grav.\  {\bf 23}, R1 (2006)

\bibitem{cl}  T.E. Clarke, P.P. Kronberg and H. B\"ohringer,  Astrophys. J.  {\bf 547}, L111 (2001).

\bibitem{cl2}  H. B\"ohringer, Rev. Mod. Astron. {\bf 8}, 295 (1995).

\bibitem{clrev} C. Carilli and G. Taylor  Ann.Rev.Astron.Astrophys.  {\bf 40}, 319 (2002). 

\bibitem{soho} See, for instance,   http://sohowww.nascom.nasa.gov/.

\bibitem{trace} See, for instance http://trace.lmsal.com/.

\bibitem{gong} See, for instance http://gong.nso.edu/.
  
\bibitem{sun} M. Dikpati, P. Gilman, and K. MacGregor,  Astrophys. J. {\bf 638}, 564 (2006); 
P.  A. Gilman and M. Miesch, Astrophys. J. {\bf 611}, 568 (2004).

\end{thebibliography}
\end{document}